\documentclass[aps,pra,pdf,superscriptaddress,amsmath,amssymb,amsfonts,twocolumn,showpacs]{revtex4-1}

\usepackage{amssymb}
\usepackage{eqnarray,amsmath}
\newcommand{\abs}[1]{\left| #1 \right|} 
\usepackage{graphicx}
\usepackage{epsfig}
\usepackage{dcolumn}
\usepackage{bm}
\usepackage{braket}
\usepackage{amsmath}
\usepackage{physics}
\usepackage{mathtools}
\usepackage{graphicx,color,xcolor}
\usepackage{hyperref}

\usepackage{hyperref}
\usepackage{multirow}
\usepackage[normalem]{ulem} %% REMOVE ME!

\begin{document}
\title{Stability and dynamics across magnetic phases of vortex-bright type excitations \\
in spinor Bose-Einstein condensates}

\author{G. C. Katsimiga}
\affiliation{Department of Mathematics and Statistics, University of Massachusetts Amherst, Amherst, MA 01003-4515, USA}
\affiliation{Department of Physics, Center for Optical Quantum Technologies, University of Hamburg, Luruper Chaussee 149, 22761 Hamburg, Germany}
\affiliation{The Hamburg Center for Ultrafast Imaging, University of Hamburg,Luruper Chaussee 149, 22761 Hamburg, Germany}	
\author{S. I. Mistakidis}
\affiliation{ITAMP,  Center for Astrophysics $|$ Harvard $\&$ Smithsonian, Cambridge, MA 02138 USA}
\affiliation{ Department of Physics, Harvard University, Cambridge, Massachusetts 02138, USA}
\author{K. Mukherjee}
\affiliation{Department of Physics, Indian Institute of Technology Kharagpur, Kharagpur, West Bengal 721302, India}
\author{P. G. Kevrekidis}
\affiliation{Department of Mathematics and Statistics, University of Massachusetts Amherst, Amherst, MA 01003-4515, USA}
\author{P. Schmelcher}
\affiliation{Department of Physics, Center for Optical Quantum Technologies, University of Hamburg, Luruper Chaussee 149, 22761 Hamburg, Germany}
\affiliation{The Hamburg Center for Ultrafast Imaging, University of Hamburg,Luruper Chaussee 149, 22761 Hamburg, Germany}

\date{\today}
	
\begin{abstract}
The static properties, i.e., existence and stability, as well as the quench-induced dynamics of vortex-bright type excitations 
in two-dimensional harmonically confined spin-1 Bose-Einstein condensates are investigated. 
Linearly stable vortex-bright-vortex and bright-vortex-bright solutions arise in both antiferromagnetic 
and ferromagnetic spinor gases upon quadratic Zeeman energy shift variations. 
Their deformations across the relevant transitions are exposed and discussed in detail evincing also that 
emergent instabilities can lead to pattern formation. 
Spatial elongations, precessional motion and spiraling of the nonlinear excitations when exposed 
to finite temperatures and upon crossing the distinct phase boundaries, via quenching of the quadratic Zeeman coefficient, are unveiled. 
Spin-mixing processes triggered by the quench lead, among others, to changes in the waveform of the ensuing configurations. 
Our findings reveal an interplay between pattern formation and spin-mixing processes being accessible in contemporary cold atom experiments.
\end{abstract}
\maketitle
	
\section{Introduction}\label{Introduction}
	
It is nowdays possible to controllably create Bose-Einstein condensates (BECs) possessing 
internal degrees-of-freedom~\cite{stamper1998optical,stenger1998spin,chang2005coherent,widera2006precision,huh2020observation}. 
These multi-component systems, due to the Zeeman splitting of the involved magnetic sublevels are known as spinor condensates 
and have been discussed in dedicated reviews~\cite{kawaguchi2012spinor,stamper2013spinor} 
and books~\cite{pethick,pitaevskii2003bose,frantzeskakis2015defocusing}. 
Among spinors with hyperfine spin $F=1$ or $2$, spin-1 BECs represent arguably the most studied class. 
The two-body interaction of spin-1 bosons features density (or interparticle) and spin-interactions. 
By engineering the internal states using optical and magnetic fields, various magnetic ground states and 
the related to them first and second order phase transitions are now accessible~\cite{kawaguchi2012spinor}. 
For instance a $^{23}$Na spinor gas experiences antiferromagnetic (AF) interactions~~\cite{stamper1998optical,stenger1998spin} 
whilst $^{87}$Rb~\cite{chang2005coherent,widera2006precision} and $^{7}$Li~\cite{huh2020observation,kim2021emission} 
feature weak and strong ferromagnetic (FM) ones.
         
The spinor ground state (GS) phase diagram has been exhaustively studied~\cite{kawaguchi2012spinor}.  
Alterations due to confinement have been only recently explored within the 
mean-field~\cite{schmied2020stability} and the many-body framework~\cite{mittal2020many}. 
Additionally, owing to the presence of internal degrees-of-freedom a plethora of nonlinear excitations 
bearing a non-topological and a topological character have been proposed theoretically. 
A partial list of the latter contains: (i) one-dimensional magnetic and unmagnetized spinor solitons~
\cite{li2005exact,zhang2007solitons,nistazakis2008bright,szankowski2011surprising,romero2019controlled,chai2020magnetic,chai2021magnetic}, 
as well as dark-antidark structures~\cite{schmied2020dark}; (ii) the realization~\cite{bersano2018three} and the ensuing phase diagram~\cite{katsimiga2021phase} of spinor dark-dark-bright and 
dark-bright-bright solitary waves, their collisions~\cite{lannig2020collisions}, as well as twisted magnetic solitons~\cite{fujimoto2019flemish};
(iii) spin domains~\cite{miesner1999observation,swislocki2012controlled}, 
monopoles~\cite{stoof2001monopoles,martikainen2002creation,dsh1}, quantum knots~\cite{dsh2}, 
as well as three-~\cite{dsh3} 
and two-dimensional (2D) 
skyrmions~\cite{marzlin2000creation,mizushima2002mermin,leanhardt2003coreless,reijnders2004rotating,choi2012observation}, 
skyrmion and meron textures~\cite{song2013ground}, non-axisymmetric vortex patterns~\cite{mizushima2002axisymmetric}.
Moreover, half-quantum vortical structures~\cite{leonhardt2000create,ruostekoski2003monopole,lovegrove2012energetically}  
can arise from the instability of singular vortices~\cite{seo2015half},
which, in turn, can emerge from the unstable dynamics of nonsingular ones~\cite{xiao2021controlled}. Filled-core vortices~\cite{sadler2006spontaneous},  
along with the very recently detected singular SO(3) vortex line~\cite{weiss2019controlled} can also be included in this list. 
It is also relevant to mention here, that the properties of specific vortex structures in 
homogeneous systems, 
such as the elliptic one characterized by 
broken axisymmetry were recently discussed 
for the polar (PO) phase in 
Refs.~\cite{PhysRevA.104.013316,takeuchi2021quantum} 
and the so-called nematic spin vortices 
appearing in the easy-plane (EP) PO phase were analyzed in 
Ref.~\cite{underwood2020properties}. 
In the same context, the robustness of confined coreless vortices when the 
longitudinal magnetization is
preserved has been 
analyzed~\cite{lovegrove2016stability}. 

\begin{figure*}
\begin{center}
\includegraphics[width=0.94\textwidth]{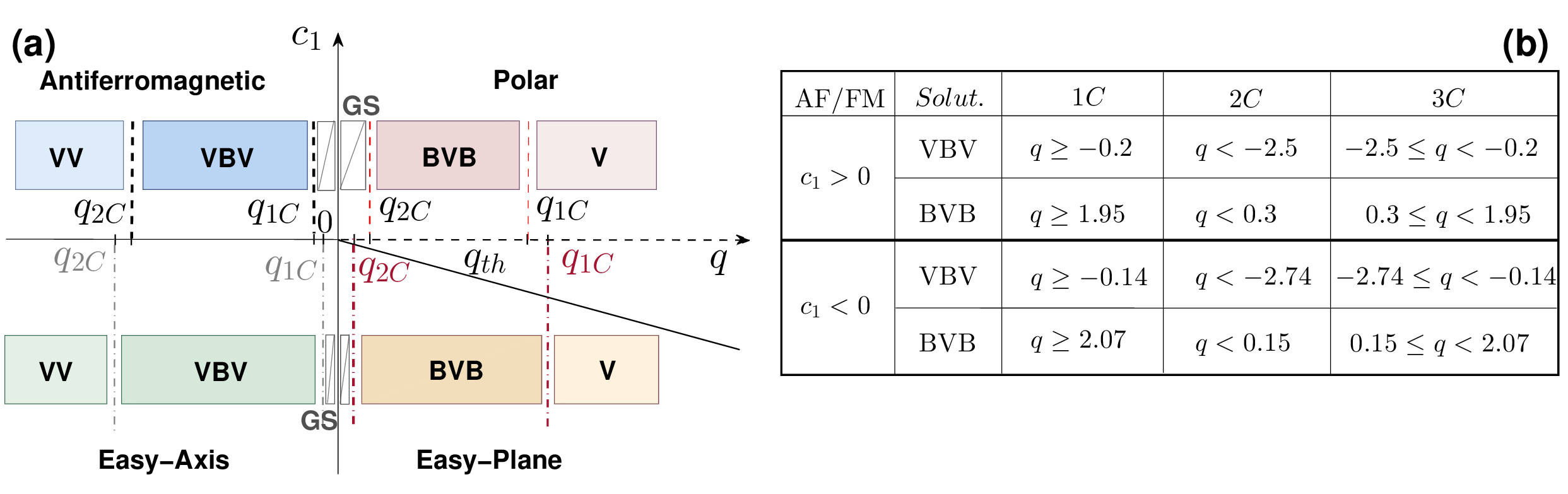}
\caption{(a) Schematic illustration of the phase diagram containing the distinct VBV and BVB stationary solutions in the $(c_1, q)$--plane as well as their corresponding deformations under QZ energy shift variations. 
(b)~Intervals of existence of nonlinear excitations of the VB type for AF, $c_1 >0$, 
(top rows) and FM, $c_1 < 0$, (bottom rows) interactions corresponding respectively to a spin-$1$ BEC consisting of  $^{23}$Na and $^{87}$Rb atoms. 
From left to right, each column depicts the occupation of a single (1C) two (2C) and all three (3C) $m_F$ components. 
A spin-component is treated as unpopulated when its occupation, $n_{m_{F}}$, is less than $1/N$. 
Here, the total number of particles $N =10^4$, while the in- and out-of-plane 
trapping frequencies are $\omega = 2 \pi \times 20$ Hz and $\omega_z = 2 \pi \times 400$ Hz respectively.}
\label{table}
\end{center}
\end{figure*}

Given the enhanced theoretical and experimental~\cite{chai2021magnetic,lannig2020collisions,chai2020magnetic,bersano2018three} 
recent interest in spinor BECs and the different excitations that can form in their distinct magnetic phases, 
we hereby consider harmonically trapped quasi-2D, spin-1 BECs featuring either AF or FM spin-interactions. 
Concerning the static properties of the two setups under consideration, we tackle spinorial stationary states 
that bear at least one vortex component being filled by bright solitons. 
In comparison to earlier studies~\cite{mizushima2002axisymmetric}, a central feature of our work is that we 
consider vortical states of the same charge and zero net magnetization. 
Also, a key property of the structures of interest herein is the filling of vortices 
with bright components when the parameters of the system permit it (see details below).
The understanding of the stability properties of such configurations, being addressed herein via a generalized 
Bogoliubov-de Gennes (BdG) theory~\cite{skryabin2000instabilities,frantzeskakis2015defocusing,kevrekidis2016solitons},  
is still far from complete. 
Only partial results of this kind exist, as e.g. in the recent study of~\cite{underwood2020properties} where  
the maximal growth rate of the so-called nematic spin vortex state is provided. 
Here, we build on earlier findings based on 
simpler one-dimensional settings~\cite{katsimiga2021phase,liu2020phase}, in order to obtain the phase 
diagram of the identified vortical states, coupled with their corresponding potential instabilities.
Triggering the latter can also be valuable, as it is strongly suggested by recent single-component BEC experiments~\cite{saint2019dynamical}, 
for designing certain topological states in the different 
spinor phases examining thereafter their dynamical response and spin-mixing processes~\cite{saint2019dynamical,PhysRevE.104.024207}.
Our findings indicate that vortex-bright-vortex (VBV) and bright-vortex-bright (BVB) excitations exist 
as stable configurations for either AF or FM spin-dependent interactions [see Fig.~\ref{table}(a)-(b)]. 
These excitations experience structural deformations upon quadratic Zeeman (QZ) energy shift variations and importantly they feature narrow QZ intervals where oscillatory instabilities occur~\cite{skryabin2000instabilities,katsimiga2020observation}. 

Dynamical evolution of perturbed VBV (BVB) entities entails, among others, their irregular
(regular) precessional motion, nucleation of cross-shaped spinor patterns, and 
potential spiraling of the ensuing waveforms. 
These are findings evincing that spinor BECs provide a fruitful platform for probing instability-related 
spontaneous pattern formation~\cite{PhysRevLett.127.113001,maity2020parametrically}. 
Further, quench-induced spin-mixing processes are unveiled under QZ energy shift variations 
at finite temperatures. 
The inclusion of thermal effects is inspired by their relevance in recent experiments~\cite{vinit2017precise,kang2017emergence}. 
Specifically, population transfer mechanisms are shown to be enhanced for larger values of the QZ coefficient and higher temperatures. 
Finally, the nonequilibrium dynamics of the vortical spinor configurations  
reveals the generic activation of their precessional motion, but also 
deformations where spinors simultaneously exhibit characteristic spatially anisotropic elongations.  

The workflow of the present effort is as follows. Section~\ref{sec:theory} 
sets up the model mean-field equations of motion and the linearization method utilized herein. 
Section~\ref{sec:results} contains our main findings regarding the existence, stability and dynamics of AF and FM spin-1 BECs. 
Their quench dynamics at finite temperatures is discussed in Sec.~\ref{quenches}. 
In Sec.~\ref{sec:conclusions} we provide a summary of our results and a list of interesting perspectives for future investigations. 
Appendix~\ref{sec:appendix}, elaborates on the impact of higher-charge vorticity  
generalizing earlier instability findings occurring in single-component settings~\cite{pu1999coherent}, 
demonstrating also dynamical triangular pattern formation. 

\section{Embedding nonlinear excitations in the spinor system}\label{sec:theory}

\subsection{Mean-Field equations}

We consider a spin-1 BEC of $N=10^4$ $^{87}$Rb~\cite{bersano2018three,klausen2001nature} or $^{23}$Na~\cite{chai2020magnetic} atoms of mass $M$.~\footnote{Notice that in such a setting phenomena associated
with multiple orbital occupation and signatures
of fragmentation should be expected to be absent;
for a relevant discussion, see, e.g.,~\cite{mittal2020many}.}
A uniform magnetic field $B$ is applied along the transversal $z$-direction, and the system is confined in a quasi-2D harmonic trap. 
The quasi-2D trap is of the form $V(x, y, z) = M \omega^2(x^2 + y^2)/2 + M \omega^2_z z^2/2$, obeying the condition $\omega_z >> \omega$. 
Here $\omega_z$ denotes the out-of-plane oscillator frequency, i.e., the one along the  $z$-direction, 
and $\omega$ refers to the frequency in the $(x-y)-$plane (alias in-plane oscillator frequency). 
The corresponding three-component wave function, 
$\vb{\Psi}(\vb{r};t)=(\Psi_1(\vb{r};t),\Psi_{0}(\vb{r};t), \Psi_{-1}(\vb{r};t))$ with $\vb{r}\equiv\{x,y,z\}$, 
represents the distinct spin-components, $m_{F} = \pm 1, 0$, of a spin-$1$ BEC. 
Additionally, throughout this work we choose as characteristic length and energy scales the in-plane 
oscillator length $l_{\rm osc} = \sqrt{\hbar/M\omega}$ and $\hbar \omega$ respectively.
Accordingly, space and time coordinates are rescaled as $x'=x/l_{\rm osc}$, $y'=y/l_{\rm osc}$, $z'=z/l_{\rm osc}$ and $t'=\omega t$ 
respectively and the wave function as $\Psi_{m_{F}}(x', y', z') = \sqrt{(l^3_{osc}/N)}\Psi_{m_{F}}(x, y, z)$. 
However, due to the quasi-2D geometry of the potential considered herein (i.e. $\omega_z >> \omega$) 
the aforementioned three-dimensional wave function can be factorized as follows $\Psi_{m_{F}}(x',y',z',t) =\Psi_{m_{F}}(x',y',t) \phi(z')$.
Here, $\phi(z')$ is the normalized GS wave function in the $z$-direction, and $\Psi_{m_{F}}(x',y',t)$ is the quasi-2D wave function.  
The latter, with the above choices and rescaling (and dropping the primes for convenience) is described within the mean-field framework 
by the following dimensionless system of three coupled Gross-Pitaevskii equations (GPE)~\cite{stamper2013spinor,bersano2018three,romero2019controlled}
\begin{equation}\label{psi_plus}
\begin{split}
i\partial_t \Psi_{1}= &\mathcal{H} \Psi_{1} + q \Psi_{1} + c_{0}(\abs{\Psi_{+1}}^2 + \abs{\Psi_{0}}^2 + \abs{\Psi_{-1}}^2)\Psi_{1}  \\&
+c_1(\abs{\Psi_{+1}}^2 + \abs{\Psi_{0}}^2 - \abs{\Psi_{-1}}^2)\Psi_{1} + c_1 \Psi^{*}_{-1} \Psi^{2}_{0},
\end{split}
\end{equation}
\begin{equation}\label{psi_zero}
\begin{split}
i\partial_t \Psi_{0}= &\mathcal{H} \Psi_{0} + c_{0}(\abs{\Psi_{+1}}^2 + \abs{\Psi_{0}}^2 + \abs{\Psi_{-1}}^2)\Psi_{0} \\&
+ c_1(\abs{\Psi_{+1}}^2 + \abs{\Psi_{0}}^2)\Psi_{0} + 2c_1 \Psi_{1} \Psi^{*}_{0}\Psi_{-1},
\end{split}
\end{equation}
	  \begin{equation}\label{psi_minus}
	  \begin{split}
	  i\partial_t \Psi_{-1}= &\mathcal{H} \Psi_{-1} + q \Psi_{-1} + c_{0}(\abs{\Psi_{+1}}^2 + 
	  \abs{\Psi_{0}}^2 + \abs{\Psi_{-1}}^2)\Psi_{-1}  \\&
	  + c_1(\abs{\Psi_{-1}}^2 + \abs{\Psi_{0}}^2 - \abs{\Psi_{1}}^2)\Psi_{-1} + 
	  c_1 \Psi^{*}_{1} \Psi^{2}_{0}.
	  \end{split}
	  \end{equation}
In the above equations, $\mathcal{H} =-\frac{1}{2}\left(\partial_x^2 +\partial_y^2\right) + V(x,y)$ 
is the single particle Hamiltonian with $V(x,y) = (x^2 + y^2)/2$ denoting the 2D harmonic potential. 
Moreover, $c_{0}$ and $c_1$ are the so-called spin-independent and spin-dependent interaction coefficients 
given by $c_{0} =\frac{2N\sqrt{2\pi \kappa}(a_{0} + 2a_2)}{3l_{\rm osc}}$ 
and $c_{1} =\frac{2N\sqrt{2\pi \kappa}(a_{2} - a_0)}{3l_{\rm osc}}$ respectively, in the units adopted herein. 
$\kappa = \omega_z/\omega$ is the anisotropy parameter, while the scattering lengths $a_{0}$ and $a_{2}$ account for collisions 
between two atoms belonging to the scattering channels with total spin $F=0$ and $F=2$ respectively. 
Additionally, $c_{0}>0$ ($c_{0}<0$) accounts for repulsive (attractive) interatomic interactions, 
while $c_1 >0$ and $c_1 < 0$ designate AF and FM spin-interactions, respectively. 
Furthermore, the QZ energy shift, $q$, can be determined via the relation $q=\mu^2_B B^2/(4 \hbar \omega E_{\rm hfs})$, 
where $\mu_{B}$ denotes the Bohr magneton and $E_{\rm hfs}$ is the hyperfine splitting. 
Notably, $q$ can be tuned experimentally either by adjusting the external magnetic 
field $B$~\cite{santos2007spinor} or the hyperfine splitting $E_{\rm hfs}$ by utilizing a microwave dressing field~\cite{leslie2009amplification,bookjans2011quantum}. 

Moreover, the total number of particles, $1\equiv \sum_{m_F} \int dx~dy~|\Psi_{m_F} (x,y,t)|^2$, 
is preserved with the population fraction of each spin component being defined as 
\begin{equation}
%n_{m_{F}} = \frac{1}{N}\int dx~dy~\abs{\Psi_{m_{F}}}^2,~~m_{F} = 0,  \pm1,
n_{m_{F}} = \int dx~dy~\abs{\Psi_{m_{F}}}^2,~~m_{F} = 0,  \pm1,
\end{equation}
and satisfying $0\leq n_{m_{F}} \le 1$. 
Throughout this work we prescribe that the (similarly conserved quantity of the) 
net magnetization along the $z$-direction i.e., $\mathcal{M}_z = \int dx~dy~\left(|\Psi_{+1}|^2 -|\Psi_{-1}|^2\right)$, remains zero. 
This, in turn, implies that there is no population imbalance between the symmetric $m_{F} = \pm 1$ components.

Below, the in-plane trapping frequency is set to $\omega = 2 \pi \times 20$ Hz and the 
transverse one to $\omega_z = 2 \pi \times 400$ Hz. 
This leads to an anisotropy parameter $\kappa=20$ inspired by recent 2D BEC experiments, see, e.g., Ref.~\cite{PhysRevLett.127.113001}. 
Additionally, for AF interactions, a BEC of $^{23}$Na atoms is considered having mass $M=23$amu, $s-$wave scattering 
lengths $a_{0} =2.52862$nm, $a_{2}=2.77196$nm and therefore, $c_{0}\approx 0.013N$ and $c_{1} \approx 0.00039N$~\cite{kawaguchi2012spinor,stamper2013spinor}. 
For FM interactions, a BEC of $^{87}$Rb atoms is employed with mass $M=87$amu, $a_{0} = 5.387$nm, $a_{2}=5.313$nm 
and thus $c_{0}\approx 0.05N$ and $c_{1} \approx -0.00023N$. 
The QZ coefficient, $q$, is typically varied within the interval $[-3, 3]$. 
The latter, has been identified to be a representative interval 
of the principal phenomenology of interest. 
Unless stated otherwise, the total particle number and the vortex charge are fixed to $N =10^4$ and $S=1$ respectively.

\subsection{Vortex-bright spinor ansatz and BdG approach} 
	 
Initially [Sec.~\ref{sec:results}], we focus on obtaining stationary solutions of the spinor system of Eqs.~(\ref{psi_plus})-(\ref{psi_minus}) 
in the form of vortex-bright (VB) solitons~\cite{law2010stable,pola2012vortex,kevrekidis2016solitons,mukherjee2020quench} 
that can occupy all three hyperfine components by utilizing a Newton-Krylov iterative scheme~\cite{kelley2003solving}. 
Specifically, in order to introduce a vortex (V) of charge $S$ and a bright (B) soliton 
in the desired $m_{F}$ component, the following ansatz is applied to the relevant wave functions
\begin{eqnarray}
\Psi^V_{m_{F}}(x,y) &=& \mathcal{H}_m(x) \mathcal{H}_n(y) 
e^{-(mx^2+ny^2)/2},\label{vortex}\\
\Psi^B_{m_{F}}(x,y) &=& \exp\big[ -(x^2+y^2)/2\big].\label{bright}
\end{eqnarray}
In Eq.~(\ref{vortex}), $\mathcal{H}_m(x)=\left(-1 \right)^m e^{x^2}\frac{d^m}{dx^m} e^{-x^2}$ 
and $\mathcal{H}_n(y)=\left(-1 \right)^n e^{y^2}\frac{d^n}{dy^n} e^{-y^2}$ are the mth- and nth-order Hermite polynomials respectively. 
A singly quantized vortex can be obtained by employing as an initial guess the $(m,n)=(1,0)$ polynomial namely 
the first excited state for the real part of the relevant wave function, and the $(m,n)=(0,1)$ for the imaginary part, respectively. 
In a similar vein, e.g. a doulby quantized vortex ($S=2$) is realized by a suitable combination of $(m,n)$ 
i.e. by using $(m,n)=(2,0)-(0,2)$ for the real part, while $(m,n)=2 (1,1)$ for the imaginary one. 
Subsequently, in sections~\ref{sec:results} and \ref{quenches}, the stability properties 
and the quench-induced dynamics of the previously identified equilibrium states are investigated. 
Notice that we restrict our investigations to the case where the components contain vortices of the same charge $S$. 
However, it would be worthwhile to consider in the future also cases in which e.g. the symmetric spin 
states include oppositely charged vortices in order to unravel the creation of patterns analogous 
to the monopoles appearing in three-dimensions. 
\begin{figure*}
\begin{center}
\includegraphics[width=0.97\textwidth]{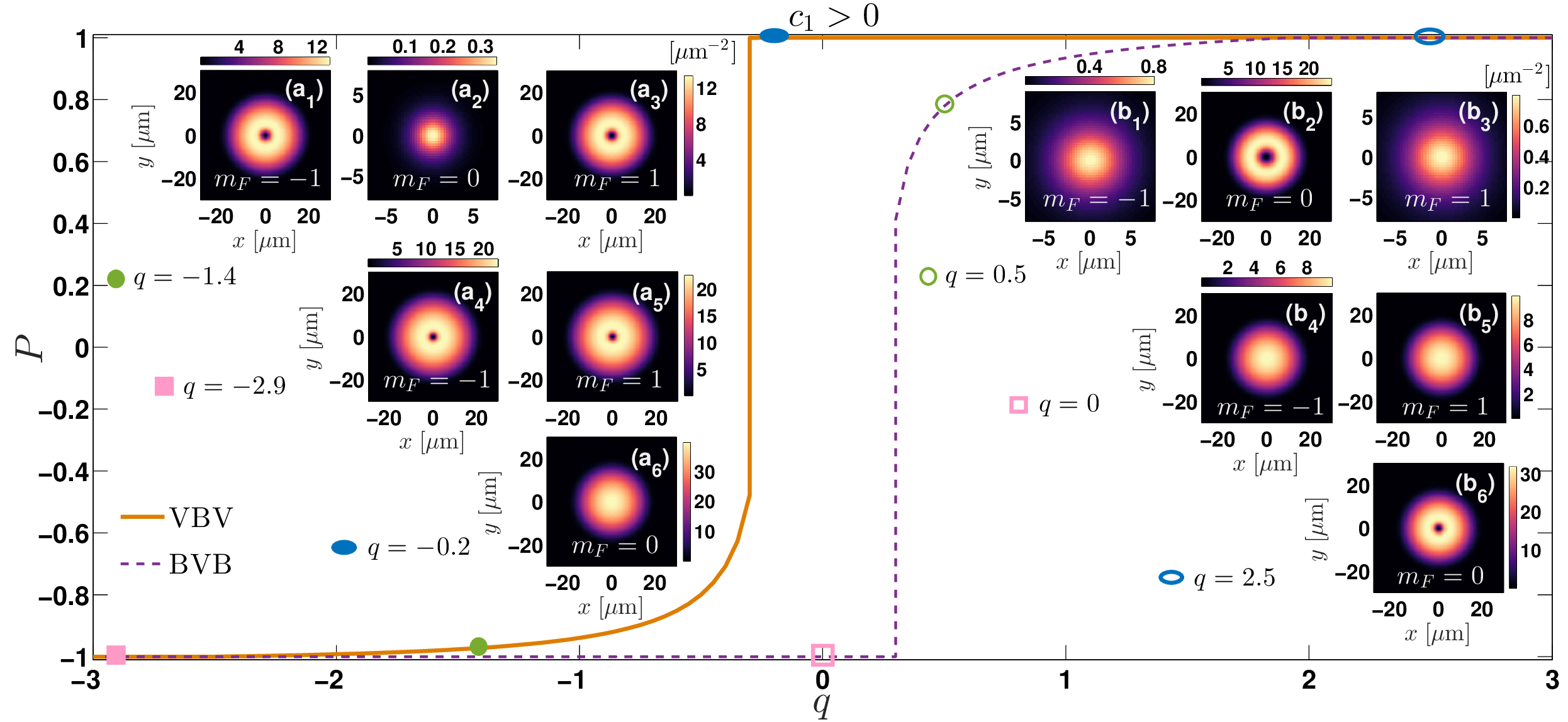}
\caption{Polarization, $P$, 
with respect to the QZ coefficient $q$ for VBV (orange solid line) and BVB (dashed purple line) 
equilibrium states existing in AF ($c_1 >0$) spin-1 BECs. 
Differently colored opaque and transparent markers indicate the value of $q$ for which the distinct solutions are provided.
Insets ($a_1$)-($a_6$) [($b_1$)-($b_6$)] illustrate representative density 
profiles, $|\Psi_{m_F}(x,y)|^2$, of a VBV [BVB] configuration (3C structure in the top row) and its corresponding 
deformations towards a 2C (middle row) and a 1C (bottom row) stationary state. 
The components that are not depicted possess zero population. 
For both types of solutions singly quantized vortices are considered for the relevant in each case $m_F$ component (see legends). 
 Notice that our results are provided in dimensionful units.}
\label{fig_1}
\end{center}
\end{figure*}

For studying the stability of the VBV and BVB configurations found herein, 
a spectral BdG analysis suitably generalized for 2D spinorial BECs is performed~\cite{skryabin2000instabilities,frantzeskakis2015defocusing,kevrekidis2016solitons,katsimiga2021phase}.
In delineating the latter, we note that it consists of perturbing the iteratively identified 
stationary states, $\Psi^0_{m_F}(x,y)$, of each phase via the ansatz
%%%
\begin{equation}
\begin{split}
\tilde{\Psi}_{m_F}(x,y,t)=\Big[\Psi^0_{m_F}(x,y)+\epsilon \Big(a_{m_F}(x,y)e^{-i\Omega t} %\nonumber 
\\ 
+ b^{\star}_{m_F}(x,y)e^{i\Omega t} \Big) \Big]  
\times e^{-i\mu_{m_F} t}. \label{BdG}
\end{split}
\end{equation}
Here, $\epsilon$ is a small amplitude perturbation parameter and $\mu_{m_F}$ with $m_F=0, \pm1$ is the chemical potential of each spin-component. 
$\Omega$ and $(a_{m_F}, b^{\star}_{m_F})^{T}$ 
denote, respectively, the eigenfrequencies and 
eigenfunctions of the resulting eigenvalue problem 
that one obtains upon substituting Eqs.~(\ref{BdG}) into the system of 
Eqs.~(\ref{psi_plus})-(\ref{psi_minus}) 
and keeping terms of order $\mathcal{O}(\epsilon)$~\cite{skryabin2000instabilities,frantzeskakis2015defocusing,kevrekidis2016solitons}. 
Namely, 
\begin{equation}
i\lambda   \left[ {\begin{array}{c}
    a_{0} \\
    b_{0} \\
    a_{1} \\
    b_{1} \\
    a_{-1} \\
    b_{-1} \\
  \end{array} } \right] =
    \left[ {\begin{array}{ccc}
    M_{1} & M_{2} & M_{3}  \\
    M_{4} & M_{5} & M_{6} \\
    M_{7} & M_{8} & M_{9} \\
  \end{array} } \right] \\
  \left[ {\begin{array}{c}
    a_{0} \\
    b_{0} \\
    a_{1} \\
    b_{1} \\
    a_{-1} \\
    b_{-1} \\
  \end{array} } \right].    
\label{eigprob}
  \end{equation}
In the above expression $\lambda \equiv -i\Omega$ and $M_j$ (with $j=1,\ldots,9$) are $2\times 2$ matrices whose explicit form is provided in Appendix~\ref{sec:appendix1}. 
The resulting eigenvalue problem of Eq.~(\ref{eigprob}) is subsequently solved numerically.
Note that in 2D spinor condensates BdG analysis of vortical configurations bearing also a bright soliton component 
is still elusive and only partial results to that effect are available, to the best
of our knowledge.

On the dynamical side, in order to study alterations of the stationary states existing in a specific 
phase when crossing a phase boundary~\cite{kiehn2019spontaneous}, a quench of the QZ energy shift is applied. 
The quench is performed from an initial (pre-quench) $q \equiv q_i$ to a final (post-quench) value $q \equiv q_{f}$ in a way that assures penetration to a different phase. 
To seed population transfer in the quench 
dynamics, the commutator of the total spin 
operator with the Hamiltonian has to be nonzero and we achieve this by including dissipation 
into the system. 
Such dissipation, can naturally arise in BEC 
experiments when 
a non-negligible thermal gas component is present in the system.
Furthermore, in the large particle limit that we operate it is expected, in line with recent 
spin-1 BEC experiments~\cite{chai2021magnetic,lannig2020collisions,chai2020magnetic,bersano2018three},  
that quantum fluctuations are suppressed. 
For the dynamical evolution of the spinorial system a fourth-order (in time) Runge-Kutta method is used 
with temporal and spatial discretization $dt=10^{-4}$ and $dx=dy=0.05$ respectively, while a (2nd order) 
finite difference scheme is utilized for the spatial derivatives.
	
\section{Static properties of VBV and BVB spinor excitations}\label{sec:results}

\subsection{Antiferromagnetic vortex-bright type configurations}

To tackle the nonlinear excitations of the VB form that arise in the distinct phases of 2D harmonically confined spin-1 BECs, 
an initial guess provided by Eqs.~(\ref{vortex})-(\ref{bright}) is introduced to 
the time-independent version of the system of Eqs.~(\ref{psi_plus})-(\ref{psi_minus}). 
Specifically, for AF interactions ($c_1>0$), it is well-known that two distinct phases 
exist depending on the value of the QZ energy shift~\cite{kawaguchi2012spinor,stamper2013spinor}. 
Namely, for $q<0$ the AF phase is realized while for $q>0$ the system resides in the PO phase. 
In the former phase and at the GS level, only the symmetric $m_F=\pm 1$ spin-components are populated. 

Thus, a natural choice for accessing the corresponding excited states is to consider an initial guess where vortices (bright solitons) 
are embedded in the $m_F=\pm 1$ hyperfine states and a bright soliton (vortex) occupies the $m_F=0$ component. 
It turns out that among these two, i.e. VBV and BVB, configurations only VBV excitations exist in the AF phase. 
Representative density profiles are illustrated as insets in Fig.~\ref{fig_1}$(\rm{a_1})-(\rm{a_3})$. 
We note in passing that for all vortex entities to be presented throughout we have 
verified that they are accompanied by the expected 
$2 \pi S$ phase winding (with $S$ denoting the 
vortex charge). 
Recall that the polarization, 
$P = \int dx~dy~\big(\abs{\Psi_{0}}^2 - \abs{\Psi_{1}}^2 -\abs{\Psi_{-1}}^2\big)$, 
is a measure of population transfer phenomena. 
It obeys $-1 \leq P \leq 1$, 
when all three $m_F$ components (3C) are populated, but $P=1$ ($P=-1$) if only the 
$m_{F} =0$ ($m_{F}=\pm 1$) state(s) is (are) populated yielding a single (two) component, 1C (2C), configuration.
The aforementioned 3C stationary states exhibit polarizations $-1<P<1$ (see orange line in Fig.~\ref{fig_1}) 
and their interval of existence is provided in the last column of Table I in Fig.~\ref{table}(b). 
VBV excitations are further found to deform upon a $q$ variation into highly localized vortices 
occupying the symmetric spin-components 
as $q$ decreases 
[Fig.~\ref{fig_1}$(\rm{a_4}), (\rm{a_5})$]. 
These 2C vortices are indeed characterized by 
$P=-1$ and they exist for all values of $q<-2.5$ that we have checked, see also 
second column of Table I in Fig.~\ref{table}(b). 
Yet another deformation occurs for the VBV configurations but upon increasing $q$. 
In this case, each vortex core gradually becomes wider in order to effectively 
trap~\cite{pola2012vortex} the accompanying wider bright soliton of the $m_F=0$ spin-component. 
This alteration holds until the 1C GS of the PO phase is reached that is, in turn, characterized 
by $P=1$ [first column of Table I in Fig.~\ref{table}(b)]. 
Notice the abrupt jump of $P$ from $P\approx -0.48$, $q=-0.3$ to $P=1$, $q=-0.2$ [blue opaque ellipse in the $P$ curve of Fig.~\ref{fig_1}] that signals the abrupt population transfer to the $m_F=0$ 1C state [Fig.~\ref{fig_1}$(\rm{a_6})$].
\begin{figure*}
\begin{center}
\includegraphics[width=0.97\textwidth]{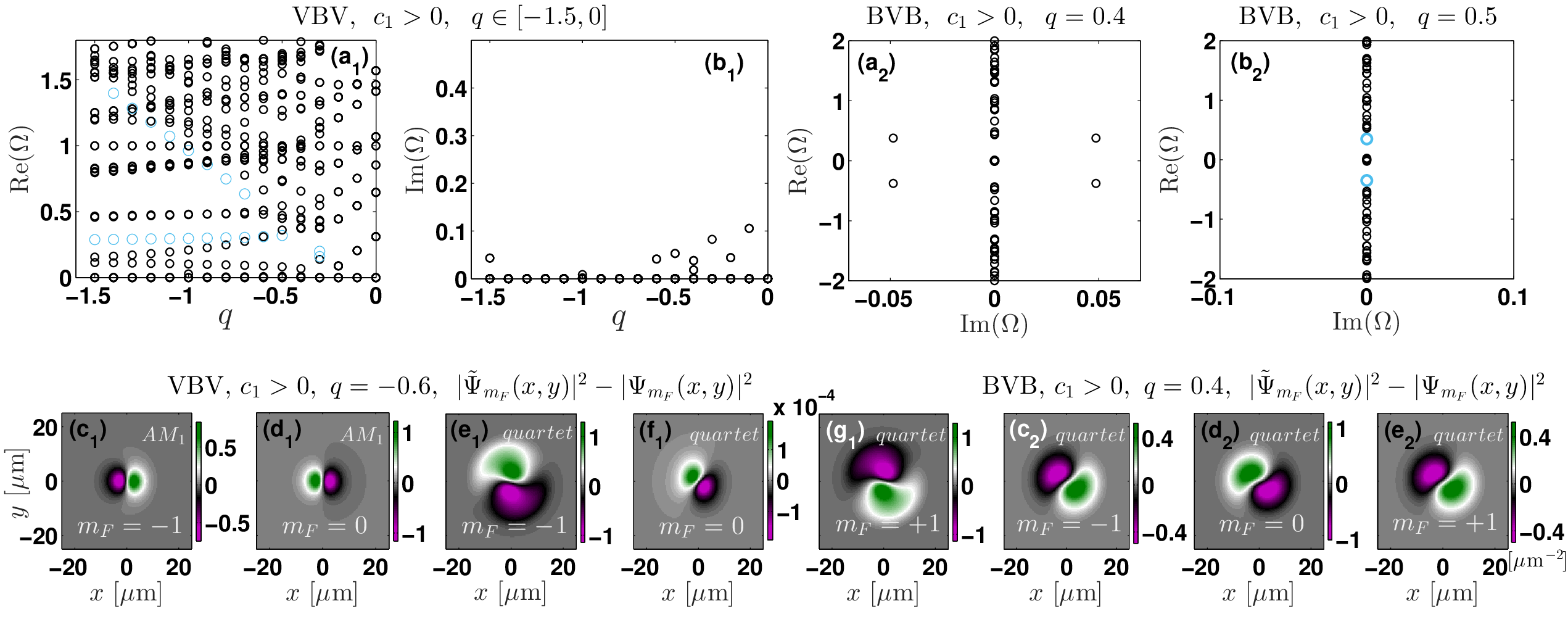}
\caption{BdG spectra of 3C $(\rm{a_1})$-$(\rm{b_1})$ VBV and $(\rm{a_2})$-$(\rm{b_2})$ BVB stationary states upon a $q$ variation for $c_1>0$.
In both cases the anomalous modes (AMs) are depicted by light blue circles while the background ones by black circles. 
Notice the two AM present for VBV structures when compared to the single pair occurring for BVB configurations. 
$(\rm{c_1})$-$(\rm{g_1})$ [$(\rm{c_2})$-$(\rm{e_2})$] 2D contour plots of the difference $\Delta \Psi_{m_F} \equiv |\tilde{\Psi}_{m_F}(x,y)|^2-|\Psi_{m_F}(x,y)|^2$, demonstrating the effect that the perturbation has on a VBV [BVB] solution for $q=-0.6$ [$q=0.4$]. 
The AF BEC consists of $N =10^4$ sodium atoms confined in a quasi-2D harmonic trap. 
Note that length and density are given in units of $[\rm{\mu m}]$ and $[\rm{\mu m}^{-2}]$ respectively.} 
\label{fig_2}
\end{center}
\end{figure*}	

Having examined the existence of VBV excitations along with their relevant structural deformations 
we next explore the stability properties of such configurations.
In contrast to earlier predictions mostly focused on energy based 
considerations~\cite{mizushima2002axisymmetric,underwood2020properties,PhysRevA.104.013316,takeuchi2021quantum} 
below we utilize a generalized BdG theory to microscopically determine the involved internal modes.
As stated earlier, to perform the BdG analysis the ansatz of Eq.~(\ref{BdG}) is used for this specific stationary solution. 
The relevant BdG spectra, obtained upon solving the eigenvalue problem of Eq.~(\ref{eigprob}) associated to the VBV solutions, 
are depicted in Fig.~\ref{fig_2}$(\rm{a_1})-(\rm{b_1})$. 
Note, that there exist in the spectrum 
three different pairs of modes lying at the $\rm{Re}(\Omega)$ axis around
the origin of the $\rm{Re}(\Omega)- \rm{Im}(\Omega)$-plane, i.e., at 
$\rm{Re}(\Omega)=\rm{Im}(\Omega)=0$. 
These zero eigenfrequencies, not visible in the scales shown, are generated by
continuous symmetries. The spinor system under study preserves the total particle number (phase invariance of the equations of motion), the magnetization and further has rotational symmetry, thus explaining the existence of these three pairs. 
Besides the aforementioned modes, two 
additional negative energy ones appear among 
the remaining modes of the discrete spectra 
that are denoted by light blue circles. 
The two distinct trajectories, obtained with respect to $q$, of these so-called anomalous modes (AMs) 
can be discerned in Fig.~\ref{fig_2}$(\rm{a_1})$. 
Each of these modes is known to correspond to the precession of each of the two vortices within the parabolic trap~\cite{law2010stable,pola2012vortex}. 
Additionally, these AMs are quantified through their negative energy or negative Krein signature~\cite{skryabin2000instabilities} which for the 2D spinor system reads  
\begin{eqnarray}
 K=\Omega~\int~dx~dy~\sum_{m_F=0, \pm 1} |a_{m_F}|^2 - |b_{m_F}|^2. \label{Krein}
\end{eqnarray}

It should be marked here that the existence of these modes is an immediate byproduct 
of the fact that the stationary states found 
herein are excited states of the spinor system. Namely, such modes would be absent in the case 
of the system's GS.
Moreover, as long as these eigenfrequencies maintain their real nature, then their negative Krein signature 
further indicates that while a stationary solution is dynamically stable, 
it is simultaneously unstable thermodynamically~\cite{frantzeskakis2015defocusing}. 
The latter, in turn, implies that given a channel of energy dissipation, as in the case of the dissipative spinor 
system that will be discussed below, these eigendirections will be activated 
leading to an instability of the ensuing configuration. 
Notice that upon increasing $q$ so as to reach the phase transition point ($q=0$), 
in the vicinity of the latter, the aforementioned negative energy modes decrease in frequency, 
with both crossing the zero frequency axis around $q\approx -0.2$. 
At the same time also a decreasing in frequency positive energy mode 
crosses $\Omega=0$ and leads to the appearance of the finite imaginary part, $\rm{Im}(\Omega)\neq 0$, shown in Fig.~\ref{fig_2}$(\rm{b_1})$. 
The destabilization of the deformed VBV configuration is followed by a change in the Krein signature of the 
two (previously) negative energy modes from negative (light blue circles) to positive (black circles). 

In addition to the above stability analysis results, there exist narrow intervals of $q$ where 
oscillatory instabilities~\cite{katsimiga2020observation} take place for the VBV solution. 
In general, this type of instability stems from collision events involving pairs of positive 
and negative Krein signature modes resulting in eigenfrequency quartets and also possessing 
a finite imaginary component $\rm{Im}(\Omega)\neq 0$~\cite{katsimiga2020observation,katsimiga2021phase}. 
We must emphasize here, that this is yet another key feature related to the 
theory of AMs: namely, their role in the manifestation of instabilities even in the absence of finite temperatures. 
Three such collision events can be readily seen in the BdG spectrum of Fig.~\ref{fig_2}$(\rm{a_1})$ 
appearing e.g. at $q=-1.5$, $q=-0.6$ and $q=-0.4$. 
The first two are associated with the higher-lying anomalous mode whose 
absence for these values of $q$ is transparent while the last one entails 
the collision and disappearance of both negative energy modes. 
\begin{figure*}
\begin{center}
\includegraphics[width=0.97\textwidth]{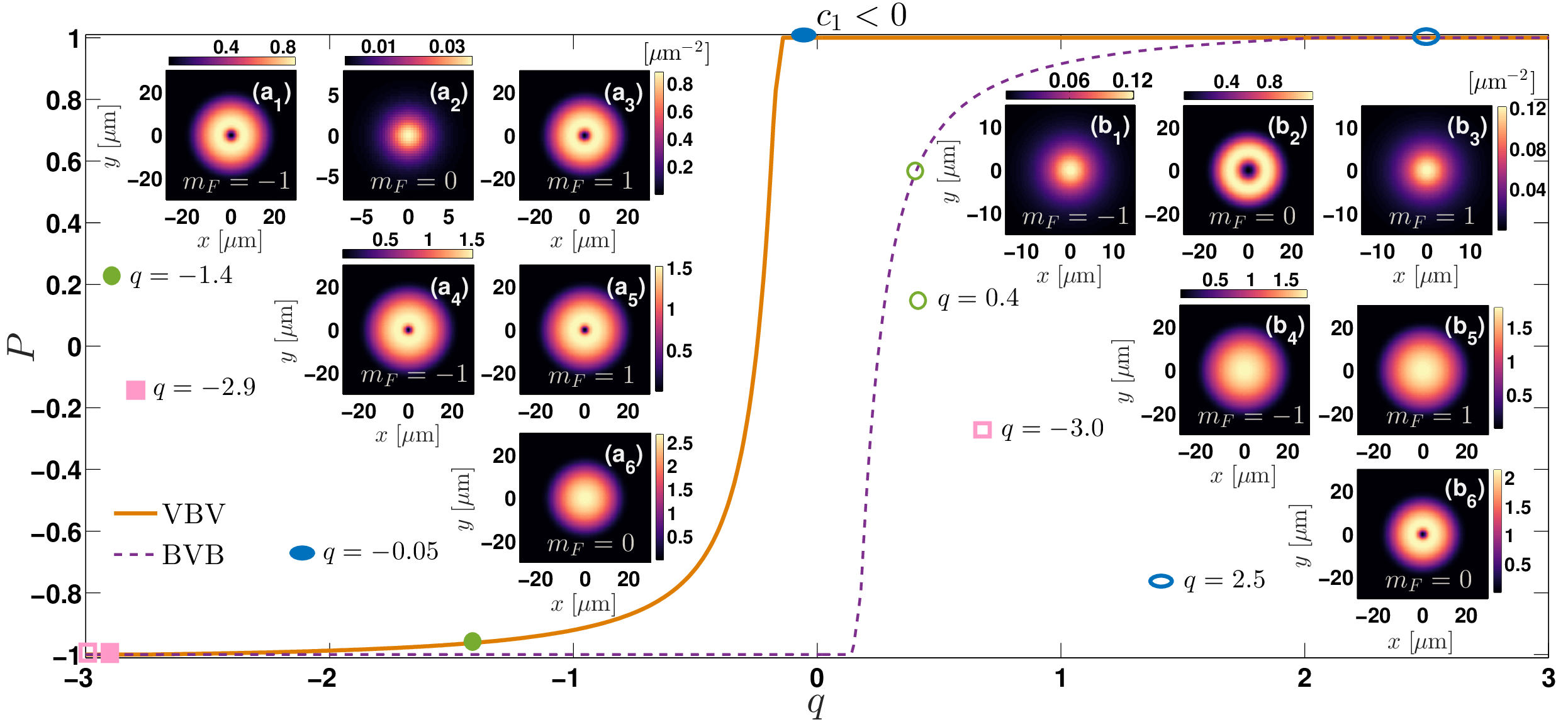}
\caption{Polarization, $P$, as a function 
of the QZ coefficient $q$ for the BVB and VBV equilibrium states occurring in the distinct phases of a spin-1 FM ($c_1 < 0$) condensate. 
Insets on the left [right] hand side ($\rm{a_1}$)-($\rm{a_6}$) [($\rm{b_1}$)-($\rm{b_6}$)] 
showcase representative examples of the $m_F$ densities, $|\Psi_{m_{F}}(x,y)|^2$, of a 3C VBV [BVB] 
configuration together with its relevant 1C and 2C deformed structures (see legends), as $q$ is varied. 
The corresponding QZ values in each of the aforementioned cases are also indicated for the individual stationary solutions,
with opaque and transparent markers in the polarization curves pointing explicitly at each specific value. Our results are presented in dimensional units, i.e. length is measured in $[\rm{\mu m}]$ and density in $[\rm{\mu m}^{-2}]$.}
\label{fig_4}
\end{center}
\end{figure*}	

Two case examples are considered below for $q=-0.6$, demonstrating the activation of e.g. the lower-lying 
anomalous mode ($AM_1$) along with exploring the oscillatory instability present for this value of $q$. 
Particularly, Fig.~\ref{fig_2}$(\rm{c_1})-(\rm{g_1})$ illustrate %density profiles
2D contours quantifying the density difference between a perturbed and an equilibrium solution
$\Delta \Psi_{m_F}(x,y) \equiv |\tilde{\Psi}_{m_F}(x,y)|^2-|\Psi_{m_F}(x,y)|^2$. 
The perturbation here, consists of adding to the VBV stationary state the eigenvector 
associated either with $AM_1$ or with the eigenfrequency quartet identified for $q=-0.6$. 
Notice the two-{\it lobe} 
structure imprinted in $\Delta 
\Psi_{m_F}(x,y)$ resembling a 2p orbital-like configuration. 
The {\it lobes} are centered around the origin of the $(x-y)-$ plane being parallel to the $y=0$ axis. They are further found to be asymmetric with respect to $x=0$ with $\Delta \Psi_{-1}(x>0,y)>0$, Fig.~\ref{fig_2}$(\rm{c_1})$ [$\Delta \Psi_{0}(x<0,y)>0$, Fig.~\ref{fig_2}$(\rm{d_1})$]. 
Moreover, the $m_F=+1$ component (not shown) has the same effect with 
that of $m_F=-1$ when the VBV is perturbed via $AM_1$ but $\Delta \Psi_{+1}(x,y)$ is complementary to $\Delta \Psi_{-1}(x,y)$ when the VBV is perturbed via the $AM_2$ mode.
However, this is not the case when considering the quartet scenario [Fig.~\ref{fig_2}$(\rm{e_1})-(\rm{g_1})$]. 
The predominant effect of 
this mode is the asymmetric distribution of $\Delta \Psi_{\pm 1}(x,y)$ with respect to $y=0$ being $\Delta \Psi_{-1}(x,y>0)>0$ [$\Delta \Psi_{+1}(x,y<0)<0$]. 
Both components are azimuthally deformed exhibiting a counterclockwise rotation.  
The $m_F=0$ one practically remains unaffected, with $\Delta \Psi_{0}(x,y)\sim 10^{-4}$ featuring an asymmetry along $x=-y$. 
Finally, it is worth commenting here, that dynamical evolution of the excited, with $AM_1$, VBV entity leads 
to its precessional motion where the entire VBV rotates around the trap. 
Whilst, exciting the configuration with $AM_2$ results in a rotating 
cross-shaped pattern in which the vortex components perform an anti-phase oscillation 
among each other and the $m_F=0$ bright component remains unaltered. 
This anti-phase vibration leads, in turn, to an overall breathing of the BEC background.  
 
For AF interactions but for $q>0$, namely within the PO phase, the preferable configuration consists of a solely occupied $m_F=0$ spin-component. 
Since this component, according to the GS of the system~\cite{stamper2013spinor}, is expected to become the majority one, 
in our search for nonlinear excitations arising in this phase we choose to imprint a vortex on it. 
Consequently, bright solitons are plugged in the remaining $m_F=\pm 1$ spin-components. 
With such an initial guess, indeed, BVB stationary solutions are captured for $0.3\le q <1.95$ 
[see also the relevant third column of Table I in Fig.~\ref{table}(b)]. 
Characteristic density contours of such a BVB structure are presented as insets in Fig.~\ref{fig_1}$(\rm{b_1})-(\rm{b_3})$. 
Notice that similarly to the VBV configurations, the BVB stationary states are characterized 
by $-1<P<1$ (see the purple line in Fig.~\ref{fig_1}) and they also experience two deformations with respect to $q$. 
One deformation is rather gradual as captured by the slope of the polarization as $q$ increases, leading to a single highly 
localized vortex occupying the $m_F=0$ hyperfine state 
[see Fig.~\ref{fig_1}$(\rm{b_6})$ and the relevant first column of Table I in Fig.~\ref{table}(b)]. 
On the contrary, as $q$ decreases towards the first order transition boundary ($q=0$) separating 
the PO and the AF phase, an abrupt deformation of the BVB configuration to the 2C state, 
reminiscent of the GS of the AF phase, occurs 
[see Fig.~\ref{fig_1}$(\rm{b_4}), (\rm{b_5})$ and the second column of Table I in Fig.~\ref{table}(b)] around $q=0.3$. 
Notice, that both the VBV and the BVB configurations feature smooth deformations towards the 2C and the 1C vortex state respectively. 
In the opposite $q$ direction a sharp transition takes place when the relevant phase boundary is approached 
to 1C and 2C zero vortex states respectively. 
This behavior of the polarization is in direct contrast to the corresponding 
sharp transition occurring on the GS level, i.e. in the absence of nonlinear excitations 
(results not shown here for brevity)~\cite{kawaguchi2012spinor,stamper2013spinor}.
\begin{figure*}
\begin{center}
\includegraphics[width=0.97\textwidth]{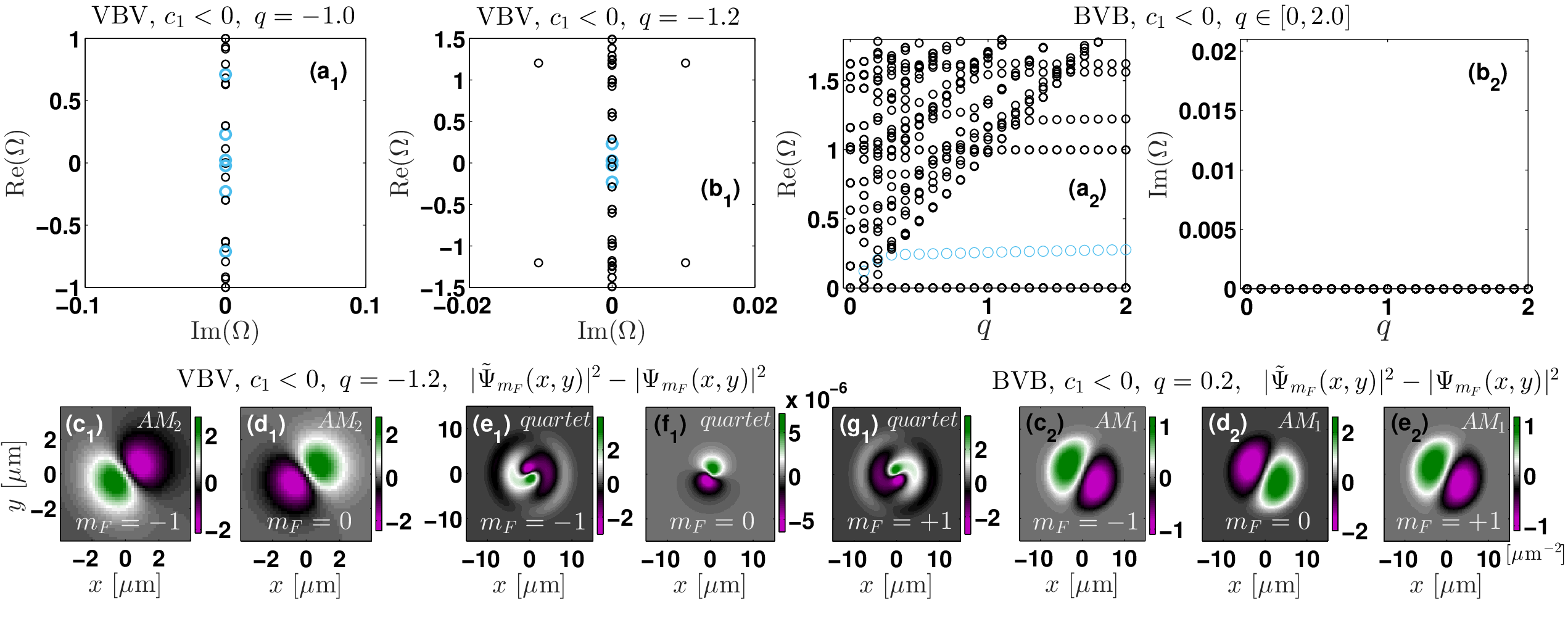}
\caption{Same as Fig.~\ref{fig_2} but for $c_1<0$. Light blue circles denote the AMs present 
in the spectra and black circles are used for the background modes. 
Contrary to AF interactions, FM VBV entities feature three anomalous mode pairs but a single pair is present for BVB configurations.
$(\rm{c_1})$-$(\rm{g_1})$ [$(\rm{c_2})$-$(\rm{e_2})$] $\Delta \Psi_{m_F} \equiv |\tilde{\Psi}_{m_F}(x,y)|^2-|\Psi_{m_F}(x,y)|^2$, quantifies the difference between a perturbed and an equilibrium VBV [BVB] solution for $q=-1.2$ [$q=0.2$].
The FM BEC consists of $N =10^4$ rubidium atoms confined in a quasi-2D harmonic trap, while length and density are given in units of
$[\rm{\mu m}]$ and $[\rm{\mu m}^{-2}]$.} 
\label{fig_5}
\end{center}
\end{figure*}	

BVB excitations turn out to be linearly stable configurations for all values of $q \in (0.4, 1.95)$, with a relevant 
example shown in the BdG spectrum of Fig.~\ref{fig_2}$(\rm{b_2})$ e.g. for $q=0.5$. 
Due to the single vortex contained in this configuration, only a single pair of negative energy modes is present in this spectrum. 
According to our discussion above,  when activated, i.e., upon adding the associated to it eigenvector to the BVB solution, 
this mode leads to the precessional motion of the BVB structure. 
It is only for significantly deformed BVB configurations, namely for states where 
the bright soliton dominates the configuration corresponding to $q\leq 0.4$, that oscillatory 
instabilities [like the one depicted in Fig.~\ref{fig_2}$(\rm{a_2})$] appear. 
In order to appreciate the effect of the emergent eigenfrequency quartet on the BVB solution, 
we have added to the latter the corresponding quartet eigenvector. 
A close inspection of the associated density difference $\Delta \Psi_{m_F}(x,y)$ 
illustrated in 
Fig.~\ref{fig_2}$(\rm{c_2})-(\rm{e_2})$, reveals that such an addition leads to 
an asymmetric across the anti-diagonal ($x=-y$) BVB structure having $\Delta \Psi_{\pm 1}(x>0,y)>0$ and $\Delta \Psi_{0}(x>0,y)<0$. 
In all cases a counterclockwise rotation takes place that is in turn related to the precessional motion of the entire BVB entity observed in the dynamics.
Finally, the anomalous mode ceases to exist for $q<0.3$ signaling the transition to the GS of the AF phase. 
Moreover, we emphasize at this point that the robustness of stable VBV and BVB stationary 
states has been also dynamically confirmed by monitoring their spatiotemporal evolution for times up to $t=2.0$s.

\subsection{Ferromagnetic VBV and BVB spinors} 

Turning to FM spin-interactions ($c_1<0$) three phases can be realized as $q$ is varied, 
supporting GS with an occupancy ranging from 1C to 3C~\cite{kawaguchi2012spinor}. 
In particular, the so-called 1C fully magnetized along the $+z$ ($-z$)-direction easy-axis (EA) phase exists for $q<0$. 
Since we operate in the regime where the harmonic oscillator length is smaller than the spin-healing length, phase separation is absent in our setting. The case where the relevant inequality is reversed, while
interesting in its own right, is outside the scope
of the present work and hence deferred to future studies. 
The 3C easy-plane (EP) phase occurs for $0<q<q_{T}$ and the 1C PO phase is  characterized by
$q\ge q_T$~\cite{kawaguchi2012spinor,katsimiga2021phase,schmied2020stability}.  
In the latter two inequalities $q_{T}=2c_1n$ (which equals $0.05$, for our chosen parameters) 
designates the threshold between the involved phases with $n$ being the peak density at the trap center. 
In this FM spinor setting, VBV stationary states are identified for $-2.74 \le q<-0.14$ [third row of Table I in Fig.~\ref{table}(b)]. 
These states possess zero net magnetization and $-1<P<1$ as shown in Fig.~\ref{fig_4}. 
They also have density profiles, $|\Psi_{m_F}(x,y)|^2$, similar to their AF  siblings [Fig.~\ref{fig_4}$(\rm{a_1})-(\rm{a_3})$]. 
Strikingly, FM VBV waves are more persistent configurations when compared to their AF counterparts. 
They are seen to penetrate deeper into the EA phase before deforming into a 2C vortex [Fig.~\ref{fig_4}$(\rm{a_4})$, $(\rm{a_5})$] structure for smaller $q$ values 
[third row of Table I in Fig.~\ref{table}(b)]. 
They further transform slower to the PO GS [Fig.~\ref{fig_4}$(\rm{a_6})$]
following an increment of $q$ towards the phase transition boundary ($q=0$). 
As such the corresponding polarization curve is found to be right-shifted thus being closer to the origin when compared to the relevant AF one. 
\begin{figure*}
\begin{center}
\includegraphics[width=0.98\textwidth]{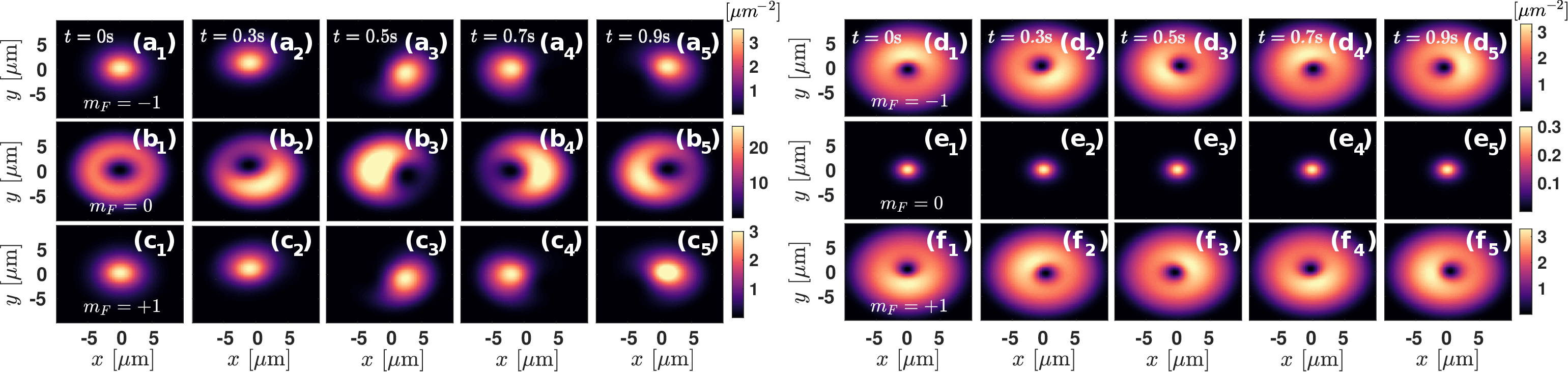}
\caption{($\rm{a_1}$)-($\rm{c_5}$) [($\rm{d_1}$)-($\rm{f_5}$)] Snapshots of the $m_F$-component 
density, $|\tilde{\Psi}_{m_{F}}(x,y)|^2$,  of a perturbed FM BVB [VBV] solution for $q=0.2$ [$q=-1.0$]. 
The distinct spin-components are illustrated respectively from top to bottom (see the legends) 
while each column corresponds to a fixed time-instant ranging from $t=0$s to $t=0.9$s. 
For both the BVB and the VBV structure the prevailing feature is their regular and irregular precessional motion, being 
activated upon adding the eigenvector associated with the single and the $AM_3$ negative energy mode respectively (see the text). All quantities shown are given in dimensionful units.}
\label{fig_5a}
\end{center}
\end{figure*}	

For $0<q<q_T$, i.e. within the EP phase, the existence of BVB stationary states is also unveiled and presented in Fig.~\ref{fig_4}. 
It is noteworthy that FM BVB structures also feature larger $q$ intervals of existence in comparison to their 
AF analogues [fourth row of Table I in Fig.~\ref{table}(b)]. 
These structures penetrate the PO regime with the underlying 3C densities as depicted in the insets of 
Fig.~\ref{fig_4}$(\rm{b_1})-(\rm{b_3})$.
Recall that at the GS level the PO phase exists for $q\ge q_T$. 
Eventually, the 3C BVB structure deforms into the 1C vortex configuration illustrated, 
e.g., for $q=2.5$ in Fig.~\ref{fig_4}$(\rm{b_6})$. 
The existence of these states is (parametrically) prolonged also following 
a decrease of $q$ until a 2C Thomas-Fermi state is reached within the EA phase [Fig.~\ref{fig_4}$(\rm{b_4})$, $(\rm{b_5})$]. 
This has as a result, a left-shifted polarization curve that is closer to the origin when compared to the relevant AF one. 

Investigating the stability of both configurations we find that, as their AF counterparts, 
VBV and BVB stationary states experience stable intervals of existence. 
This result can be verified by inspecting the BdG spectra shown in Fig.~\ref{fig_5}$(\rm{a_1})$ 
for the VBV solution and in Fig.~\ref{fig_5}$(\rm{a_2})$ and $(\rm{b_2})$ for the BVB one. 
Notice that in both cases and for the parametric intervals shown, 
all eigenfrequencies maintain their real nature, i.e., $\Im(\Omega)=0$. 
However, these structures further feature narrow $q$ intervals where oscillatory instabilities occur. 
One such example is presented regarding the VBV entity for $q=-1.2$ in the BdG spectrum of Fig.~\ref{fig_5}$(\rm{b_1})$. 
Similarly to the AF cases discussed above, also here the emergence of an eigenfrequency quartet is observed, that owes its 
presence to the collision of the higher-lying negative energy mode, $AM_3$, with a positive energy one. 
Importantly though, and also in sharp contrast to the AF VBV solutions, 
three instead of two AMs appear in the spectrum of this configuration. 
As stated earlier, since two vortices participate in this configuration 
two anomalous mode pairs are to be expected for this stationary state. 
Thus, we initially investigate further the presence of the lowest-lying mode, namely $AM_1$. 
This mode appears remarkably close to the zero eigenfrequency axis and remains near the latter as $q$ is varied till 
its destabilization slightly below the threshold separating the EA and the EP, i.e. at $q=-0.2$. 
$\Delta \Psi_{\pm 1}(x,y)$
has a four {\it lobe} spatial distribution 
closely resembling a $3d_{xy}$ orbital 
configuration which is further found to be slightly rotated counterclockwise with respect to the $x=0$ axis. 
A similar outcome is evidenced for $S=2$ VBV solutions as discussed in Appendix~\ref{sec:appendix} and visualized e.g in Fig.~\ref{fig_7} $(\rm{h_1})$.
On the other hand, $AM_1$ has a vanishing impact on the relevant
bright component, with 
$\Delta \Psi_{0}(x,y)\sim 10^{-8}$.
Particularly, $AM_1$ leads dynamically to an anisotropic spatial elongation 
of the two vortices that perform a precessional type of motion but 
with the vortices in the $m_F=\pm 1$ components rotating with a $\pi$ phase 
difference among each other and a bright soliton that remains put throughout the evolution. 
As such, this is a mode involving inter-component dynamics, rather than the intra-component ones, associated 
with the vorticity of the VBV structure.

Next, we appreciate the effect that the remaining two AMs have on VBV solutions while we note that their 
destabilization takes place at $q=-0.05$. 
Considering the eigenvector related to $AM_2$ results in an asymmetric 2p orbital-like  distribution of $\Delta \Psi_{m_F}(x,y)$, with the two {\it lobes} oriented along the anti-diagonal $x=-y$ as showcased in 
Fig.~\ref{fig_5}$(\rm{c_1})$, $(\rm{d_1})$.
It also holds that $\Delta \Psi_{-1}(x>0,y)<0$ and $\Delta \Psi_{0}(x>0,y)>0$. 
Note that a similar 2p orbital configuration is also obtained for FM $S=2$ VBV spinors (see the relevant discussion around $AM_3$ 
and $AM_4$ in Appendix~\ref{sec:appendix}).  
This mode leads upon activation to the normal or regular precession of the VBV structure.
Namely, the two vortices are on the same side and oscillate around the trap center with the bright soliton following their motion. 
A much more drastic deformation is evidenced when the solution is perturbed through 
the eigenvector of $AM_3$ leading to an asymmetric azimuthally rotated $\Delta \Psi_{m_F}(x,y)$ for the symmetric 
$m_F=\pm 1$ vortex components analogous to the one found for AF VBV equilibrium states [see Fig.~\ref{fig_2}$(\rm{e_1})$, $(\rm{g_1})$].
Also here, $\Delta \Psi_{0}(x,y)\sim 10^{-6}$
has a vanishing effect for the bright soliton
component.
As we shall show in the dynamics below, once excited, the mode $AM_3$ leads to a different form of precession of the VBV solution. 
Here, the precession of the VBV consists of two vortices hosted in the $m_F=\pm 1$ being anti-diametrically located with 
respect to the center and performing oscillations that have a $\pi$ phase 
difference with respect to one another, while 
the $m_F=0$ bright soliton component remains 
intact. 
However, this motion becomes responsible for an instability when $AM_3$ collides with a positive Krein background mode. 
Recall that whenever such a collision takes place an eigenfrequency quartet occurs in the BdG spectrum instead of the ensuing $AM$ pair. 
Indeed, notice that e.g. $AM_3$ is absent in Fig.~\ref{fig_5}$(\rm{b_1})$ giving rise to the observed quartet. 
In this latter case as it is shown in 
Fig.~\ref{fig_5}$(\rm{e_1})$-$(\rm{g_1})$, a 
spiral is imprinted in the density difference 
$\Delta \Psi_{\pm 1}(x,y)$ being of a complementing nature among these two hyperfine components, yet minuscule for the $m_F=0$ one [Fig.~\ref{fig_5}$(\rm{f_1})$]. 
This leads in turn dynamically, to a spiraling of the 2D VBV entity, an outcome caused by the oscillatory instability.    
\begin{figure*}
\begin{center}
\includegraphics[width=0.85\textwidth]{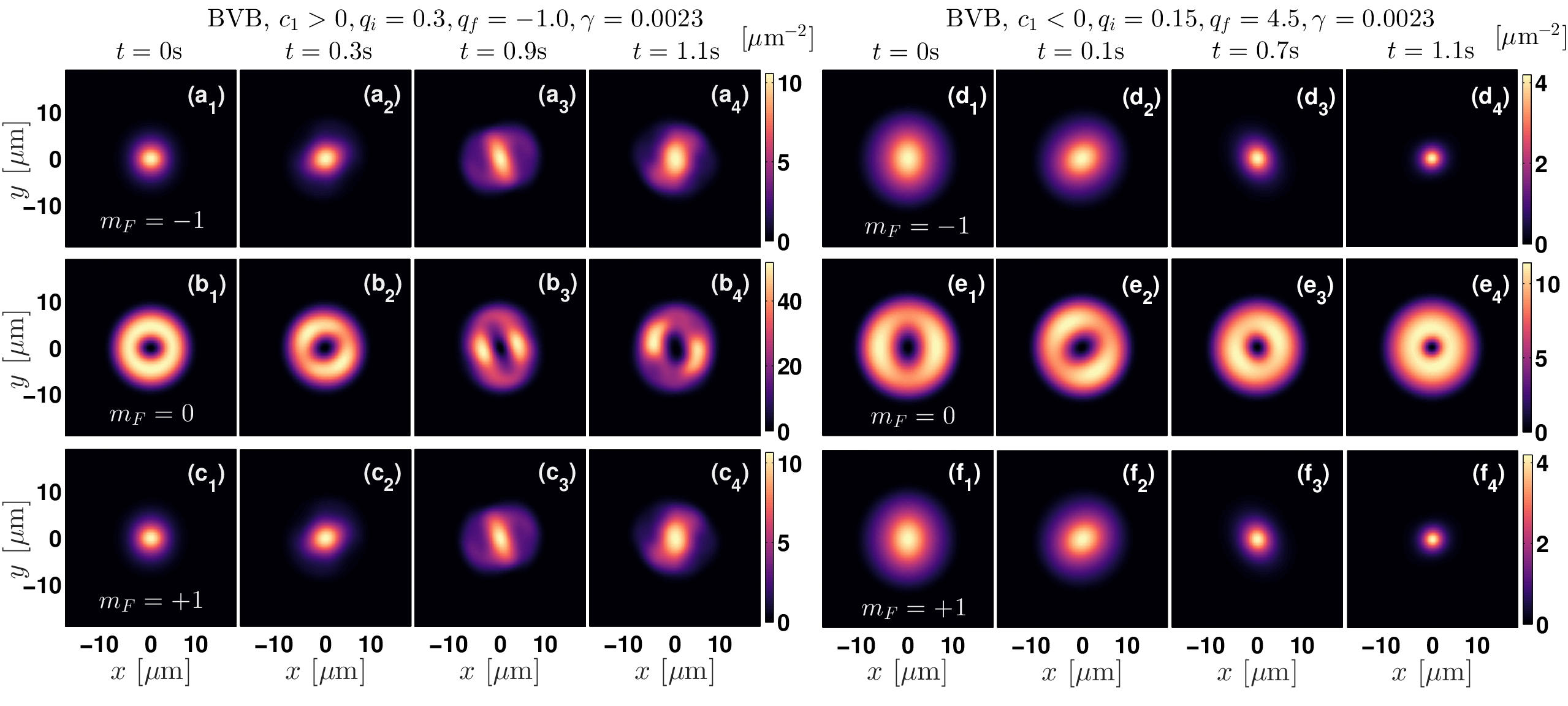}
\caption{Instantaneous density profiles of a BVB solution illustrating the $m_{F}=-1$ 
(top), $m_{F}=0$ (middle) and $m_{F}=+1$ (bottom) components, upon considering quenches 
that: $(\rm{a_1})$--$(\rm{c_4})$ either cross the phase boundary separating the PO and the AF phase of an AF spin-1 BEC
or $(\rm{d_1})$--$(\rm{f_4})$ enter deeper in the PO regime for a FM spinor gas respectively (see legends). 
For AF interactions, the structural deformation of the BVB entity corresponding to a precession and 
simultaneous spatial elongation of all components is monitored for times up to $t=1.1$s. 
The vortical pattern at $m_F=0$ acquires a dipolar spatial form. 
In contrast, the precessional motion of the BVB configuration along with a simultaneous population 
transfer from the $m_F=\pm 1$ states to the $m_F=0$ component dominates the evolution for FM interactions. 
In both cases the damping parameter is $\gamma=0.0023$ while each system contains $N = 10^4$ atoms. Note that time, length and density are measured in units of ${[\rm s}]$, ${[\rm \mu m}]$, and ${[\rm \mu m}^{-2}]$ respectively.}
\label{fig_8}
\end{center}
\end{figure*}	

As an example for the BVB solution, we choose the one of a significantly deformed, i.e., 
close to threshold, BVB excitation [Fig.~\ref{fig_5}$(\rm{c_2})$-$(\rm{e_2})$]. 
It turns out that, the bright soliton hosted in the $m_F=\pm 1$ spin-components dominates the configuration for $q=0.2$. 
This bright dominated entity is additionally found to be significantly broadened. 
Its width becomes comparable to the size of the background cloud, suggesting that the BVB character of this solution is lost. 
Perturbing this state with the eigenvector associated with the single ---in this case--- AM pair, 
leads to a two {\it lobe} 
asymmetric density difference resembling a 2p orbital for all three hyperfine states. 
The two {\it lobes} are oriented along the diagonal but experience an asymmetry, with $\Delta \Psi_{\pm 1}(x,y)<0$ for $x=-y$ and $\Delta \Psi_{0}(x,y)>0$, along the anti-diagonal. 
Featuring in this way, a similar yet inverted behavior to the one found for FM VBVs but also to FM $S=2$ BVB spinors (Appendix~\ref{sec:appendix}) when perturbed by Snapshots during the spatiotemporal evolution of this perturbed entity are provided in Fig.~\ref{fig_5a}$(\rm{a_1})-(\rm{c_5})$. 
As expected, the precessional motion of the entire BVB structure is observed 
from the initial stages of the dynamics, with the bright soliton $m_F=\pm 1$ components 
remaining trapped in the course of the evolution around the vortex core, see Fig.~\ref{fig_5a}$(\rm{a_1})-(\rm{c_5})$. 
For comparison, in the bottom panels of Fig.~\ref{fig_5a}$(\rm{d_1})-(\rm{f_5})$, a perturbed 
VBV excitation via the eigenvector of $AM_3$ is presented for $q=-1.0$. Two key findings are worth commenting here. 
The one concerns the fact that even though the amplitude of the perturbation for both structures is the same, 
the precession of the VBV excitation is not as pronounced as the one observed for the deformed BVB solution. 
However, and even more importantly irregular precession is featured by the VBV 
structure with the two vortices being out-of-phase throughout their motion. 
This is an outcome that has a drastic effect also on the bright 
soliton which, contrary to the BVB state, now remains unaffected.

Finally, in order to emulate the presence of a finite thermal fraction being usually present 
in cold atom experiments we introduced the following ansatz 
$\Psi^{\rm{pert}}_{m_F}=\Psi^0_{m_F}(x,y)\left[1+\varepsilon \delta(x,y) \right]$ 
to the $m_F$ component wave function~\cite{proukakis2008finite}. 
In this expression, $\epsilon$ accounts for the thermal fraction and $\delta(x,y)$ denotes a normally distributed 
perturbation with zero mean and variance unity~\cite{PhysRevLett.127.113001}. 
Generically, this ansatz allows for the activation of the respective AM in the course of the evolution. 
Additionally, it should be noted that the AMs are converted to unstable eigendirections in the presence of a thermal 
fraction, correspondingly dominating the BEC dynamics, similarly to what is known, e.g., for two-component
condensates~\cite{Achilleos_2012}.  
This way, the destabilization mechanisms found above would be evident in a corresponding experimental realization.
\begin{figure*}
\begin{center}
\includegraphics[width=0.9\textwidth]{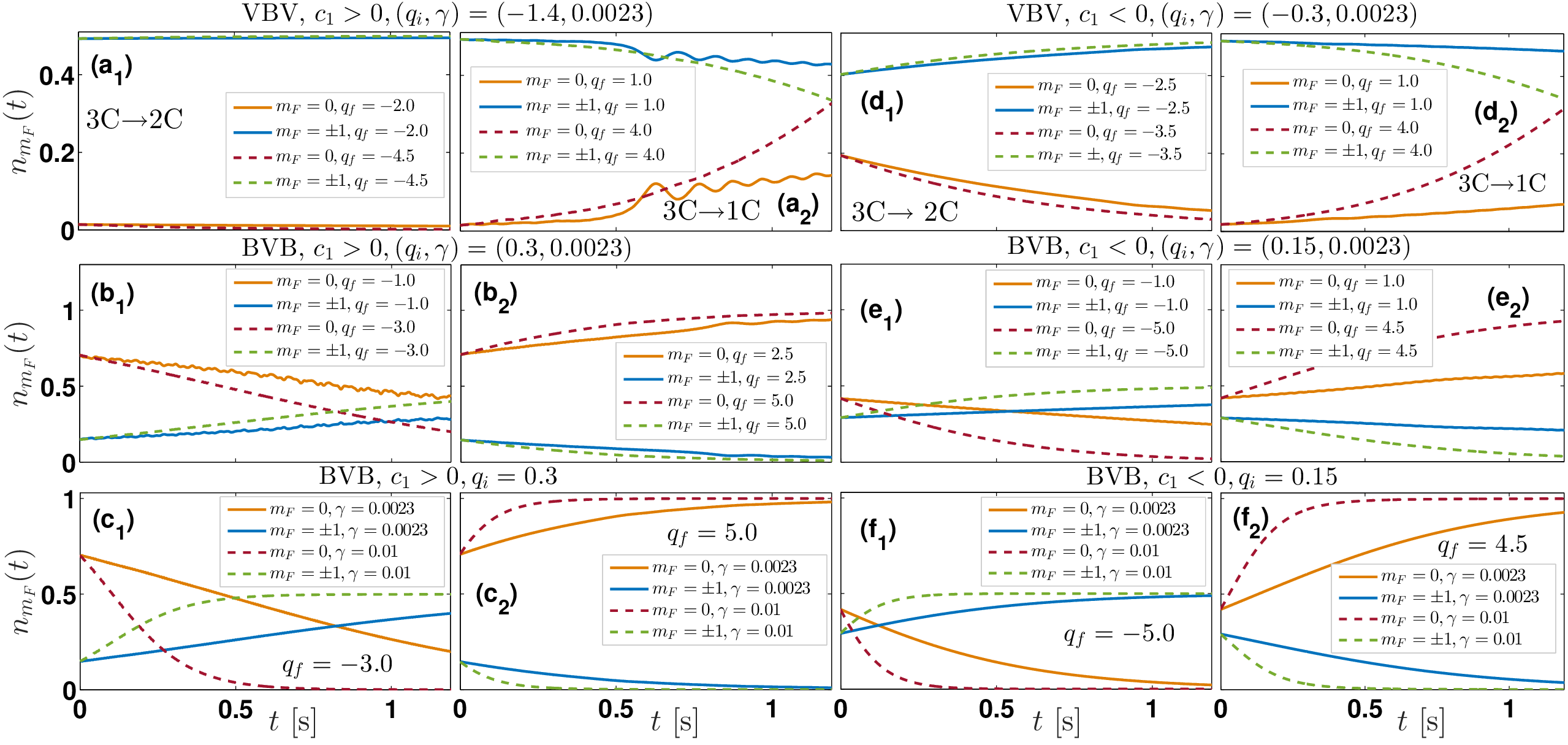}
\caption{ $(\rm{a_j})$--$(\rm{c_j})$ [$(\rm{d_j})$--$(\rm{f_j})$] with $j=1,2$, temporal evolution of the populations, $n_{m_{F}}(t)$, 
of the different spin-components considering quenches of the QZ coefficient at finite temperatures both within the same 
magnetic phase and upon crossing distinct phases of an AF [FM] spinor gas (see legends). 
Spin-mixing is triggered in all cases being more suppressed for VBV excitations as compared to the BVB ones and for both types of entities when the 
relevant transition entails quenches within the same phase. 
In all cases, the initial state configuration, having a pre-quench value $q\equiv q_i$, refers to the underlying in each 
phase 3C stationary state transitioning either from 3C$\rightarrow$2C or from 3C$\rightarrow$1C states being characterized by 
different post-quench QZ coefficients $q\equiv q_{f}$ (see legends). 
All quenches are considered for fixed $\gamma$ that is either $\gamma=0.0023$ or $\gamma =0.01$ (see legends). 
The AF [FM] condensate $c_1 > 0$  [$c_1 < 0$] consists of $N = 10^4$ $^{23}$Na [$^{87}$Rb] atoms. Note also that time is provided in dimensional units being of the order of few seconds.}
\label{fig_10}
\end{center}
\end{figure*}	

\section{Quench dynamics across magnetic phases}\label{quenches}

Having explicated the static properties of VBV and BVB nonlinear excitations, in the following we aim at addressing 
alterations of the ensuing waveforms being subjected to quenches of the $q$ parameter in order to cross the distinct magnetic phase boundaries 
(see also Fig.~\ref{fig_1} and Fig.~\ref{fig_4}). 
To monitor the quench-induced dynamical evolution of the spinor gases at hand in an experimentally relevant 
fashion~\cite{huh2020observation}, we expose them to finite temperatures. 
Note that quenches are routinely utilized in spin-1 ultracold atom experiments to probe transition boundaries~\cite{vinit2017precise}, 
spin-turbulence and the related to it half-quantum vortex generation~\cite{kang2017emergence} but also to study matter-wave jet formation~\cite{kim2021emission}. 
Contrary to the above, here we use quenches at finite temperatures i) to activate the internal motion of the identified vortical spinors, ii) facilitate population transfer among the components and iii) study structural deformations of both BVB and VBV configurations across the distinct magnetic phases.
In the mean-field framework in order to qualitatively account for thermal effects we utilize the following coupled system 
of three dissipative GPEs~\cite{proukakis2008finite,katsimiga2021phase}
\begin{eqnarray}
\left(i-\gamma \right) \partial_t \Psi_{0}&=& \mathcal{\tilde{H}} \Psi_{0} + c_{0}(\abs{\Psi_{+1}}^2 + \abs{\Psi_{0}}^2 + \abs{\Psi_{-1}}^2)\Psi_{0} \nonumber \\
&+& c_1(\abs{\Psi_{+1}}^2 + \abs{\Psi_{0}}^2)\Psi_{0} + 2c_1 \Psi_{1} \Psi^{*}_{0}\Psi_{-1}, \nonumber \\
\label{psi_zerodis}\\
\left(i-\gamma \right)\partial_t \Psi_{\pm 1}&=& \mathcal{\tilde{H}} \Psi_{\pm 1} + c_{0}(\abs{\Psi_{+1}}^2 + \abs{\Psi_{0}}^2 + \abs{\Psi_{-1}}^2)\Psi_{\pm 1} \nonumber \\
&+& c_1(\abs{\Psi_{\pm 1}}^2 + \abs{\Psi_{0}}^2 - \abs{\Psi_{\mp 1}}^2)\Psi_{\pm 1} \nonumber \\
&+& q \Psi_{\pm 1} + c_1 \Psi^{*}_{\mp 1} \Psi^{2}_{0}, \label{psi_pmdis}  
\end{eqnarray}
In Eqs.~(\ref{psi_zerodis})-(\ref{psi_pmdis}) $\mathcal{\tilde{H}} \equiv \mathcal{H}-\mu_{m_F}$ and $\gamma \ll 1$ 
is a dimensionless dissipative parameter that is connected to the spinor systems' temperature~\cite{yan2014exploring}. 
Typically, $\gamma \in [2 \times 10^{-4}, 2 \times 10^{-3}]$ refers to temperatures $T \in [10, 100]$nK as has been discussed, e.g., 
in Ref.~\cite{yan2014exploring}. 
	
Representative examples among the extensive investigations performed herein, are presented in Fig.~\ref{fig_8}$(\rm{a_1})$--$(\rm{c_4})$ 
and Fig.~\ref{fig_8}$(\rm{d_1})$--$(\rm{f_4})$ regarding the density evolution for AF and FM spin-interactions respectively with $\gamma=0.0023$. 
In the former case, we monitor the dynamics of an AF BVB excitation once quenched from the PO phase having $q_i=0.3$ 
towards the AF phase with postquench QZ coefficient $q_f=-1.0$. 
It becomes apparent that population transfer from the $m_F=0$ to the $m_F=\pm1$ states 
takes place [see also Fig.~\ref{fig_10}$(\rm{b_1})$] from the initial stages of the quench-induced dynamics triggering 
the precession of an initially stationary spinorial BVB structure. 
This motion is accompanied by a prominent elongation along with the instantaneous rotation of all three spin constituents. 
Moreover, the vortex experiences a structural deformation reminiscent of a doughnut-like pattern: an 
outcome that is further captured by the two mode motion of the relevant temporal evolution of 
the populations of the individual components illustrated in Fig.~\ref{fig_10}$(\rm{b_1})$. 
This two mode motion is characterized by rapid oscillations of the populations and a long-time transfer 
(not shown in the presented timescales) where exchange of the populations between the $m_F=0$ and $m_F=\pm 1$ takes place. 
Notice that the bright soliton $m_F=\pm 1$ components remain trapped around the vortex core, following its composite motion throughout the evolution. 
Turning to FM interactions and upon considering a quench from $q_i=0.15$ (EP phase) to $q_f=4.5$ (PO phase) 
it is observed that the precessional motion constitutes the dominant dynamical mode, entailing an arguably 
faster spin-mixing process when compared to the aforementioned AF scenario.  

In order to shed light onto the underlying spin-mixing processes triggered by the quench, 
a close inspection of the temporal evolution of the population of the individual components, $n_{m_F}(t)$, is performed. 
Specifically, Fig.~\ref{fig_10}$(\rm{a_1})-(\rm{c_2})$ and Fig.~\ref{fig_10}$(\rm{d_1})-(\rm{f_2})$ 
capture the essence of our findings for a wide selection of pre- and post-quench 
QZ energies and for distinct $\gamma$ values. 
AF ($c_1>0$) and FM ($c_1<0$) condensates are treated on equal footing. 
For both spinor settings, transitions across the distinct magnetic phases are initiated from the relevant 
in each phase 3C VBV and BVB stationary states towards the corresponding 2C or 1C configuration. 

Particularly, our key observations are the following.
Irrespectively of the spinorial BEC system, spin-mixing processes are activated from the initial stages of the quench-induced dynamics. 
We find that population transfer occurs faster for larger post-quench values $q_f$ accessing this way states that are deeper in 
the relevant magnetic phase [Fig.~\ref{fig_10}$(\rm{a_1}), (\rm{b_2})$ and Fig.~\ref{fig_10}$(\rm{d_1}), (\rm{e_2})$]. 
However, it is found to be more suppressed for VBV excitations as compared to BVB ones. 
This suppression occurs also for both types of 
entities when the relevant transition entails quenches within the same phase when compared to transitions that cross distinct phase boundaries.
Additionally, spin-mixing is accelerated for a larger dissipation parameter $\gamma$ being in turn related to higher temperatures, 
see for instance Fig.~\ref{fig_10}$(\rm{c_1}), (\rm{c_2})$ and Fig.~\ref{fig_10}$(\rm{f_1}), (\rm{f_2})$. 
We also remark that slightly enhanced intercomponent population transfer arises for AF rather than FM interactions 
as can be inferred by comparing Fig.~\ref{fig_10}$(\rm{a_2})$ and Fig.~\ref{fig_10}$(\rm{d_2})$ due to the larger spin-spin interaction in the former case.
Finally, it is important to note here, that similar to the aforementioned findings occur 
during the nonequilibrium dynamics of higher charge excitations. 
However, in this case, the spin-mixing processes discussed above, are found to be relatively accelerated.
	
\section{Conclusions and future perspectives}\label{sec:conclusions}
	
In the present work the existence, stability as well as the quench-induced dynamics of 
VB-type nonlinear excitations arising in 2D harmonically trapped spin-1 antiferromagnetic and ferromagnetic BECs have been explored. 
Our investigation has been focusing on variations of the quadratic Zeeman energy shift so as to access 
and subsequently cross the distinct magnetic phases of such settings. 
A systematic Bogoliubov de-Gennes linearization analysis has been utilized for 
the extraction of the stability properties of the considered nonlinear excitations.

In particular, the existence of VBV and BVB stationary states has been exemplified, with the former being present 
in the antiferromagnetic and the easy-plane phases for antiferromagnetic and ferromagnetic spin-interactions respectively. 
On the contrary, BVB solutions appear in the polar phase of either antiferromagnetic or ferromagnetic spinors. 
In this latter scenario, stable BVB structures are also found within the easy-plane phase. 
In both settings deformations of the ensuing waveforms as the associated transition boundary is approached are explicated 
complementing this way the phase diagram of this type of nonlinear excitations in the $(c_1,q)-$plane. 

It turns out that independently of their flavor and also of their charge, the aforementioned 
entities exhibit stable intervals of existence that can be interrupted by narrow windows where oscillatory instabilities take place. 
Indeed, we have elaborated on the number of anomalous mode eigendirections that the structures bear and thus the number of potential instabilities, 
as well as illustrated when these instabilities may materialize as a result of collision of these anomalous modes with positive energy ones. 
We have also monitored the dynamical outcome of excitation of the different anomalous modes. 
The robustness or unstable dynamics of the above-described entities are 
confirmed accordingly, demonstrating for instance the precessional motion of VBV and BVB spinors and 
their structural deformation towards ---among others--- triangular-shaped patterns.

We have further investigated the quench-induced dynamical evolution of the aforementioned three-component spinors at finite temperatures 
so as to appreciate the system's dynamical response. 
Here, it is found that spin-mixing processes occur faster for larger 
postquench quadratic Zeeman energy shifts and an increasing dissipation parameter. 
Also, population transfer is slightly enhanced when considering antiferromagnetic instead of ferromagnetic spin-dependent interactions. 
Monitoring the nonequilibrium dynamics reveals, among others, the activation of the precessional motion along 
with a spatial elongation of the spinorial nonlinear excitations, irrespectively of their specific nature and spin-interactions. 
The above processes are accelerated when higher charge vortices are contained in the spinorial configuration. The latter also bear 
a significantly larger number of anomalous modes and, thus, potentially unstable eigendirections.
\begin{figure*}
\begin{center}
\includegraphics[width=0.9\textwidth]{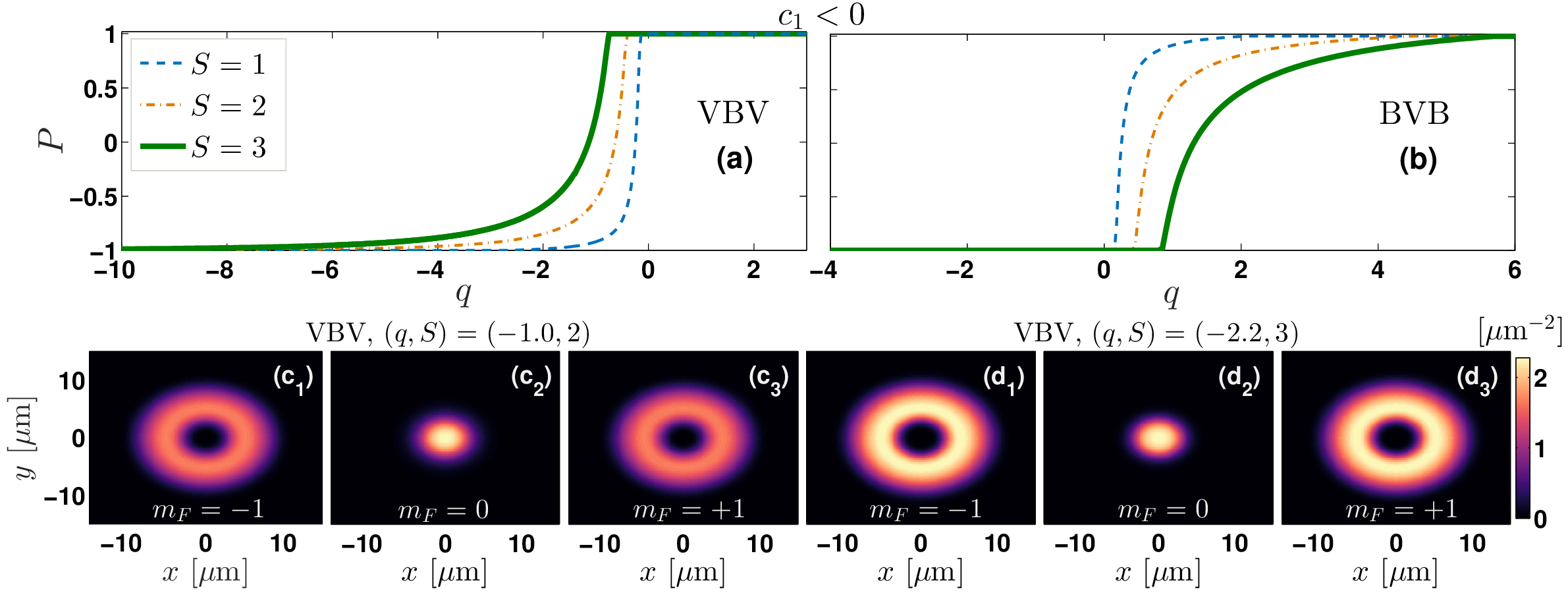}
\caption{Polarization, $P$, in terms of the QZ coefficient $q$ 
for (a) a VBV and (b) a BVB configuration upon also varying the vortex charge $S$ (see legend). 
An increasing $S$ prolongs the region of existence of the 3C state with respect to $q$. 
$(\rm{c_1})-(\rm{c_3})$ [$(\rm{d_1})-(\rm{d_3})$] Density contours of a stationary VBV state 
of charge $S=2$ [$S=3$] for $q=-1.0$ [$q=-2.2$], i.e. within the EA phase. 
The FM spin-1 BEC mixture contains $N=10^4$ $^{87}$Rb atoms.}
\label{fig_6}
\end{center}
\end{figure*}	

There exist several extensions of the present work worth pursuing in future endeavors. 
A straightforward generalization would be to study the quench dynamics in a $^7$Li spin-1 BEC where the strong 
ferromagnetic spin-interaction would certainly enhance the spin-mixing processes which might be possibly associated with a richer pattern formation. 
A detailed investigation of e.g. $S=3$ vortical spinors, that we barely touched upon herein, in symmetry broken settings would facilitate the engineering of exotic pattern formation with atomic orbital-like signatures.
Additionally, exploring the interaction effects of vortex lattices as well as their stability and 
dynamics in spinor setups is of direct relevance, due to the potential of inclusion of external rotation~\cite{pethick,pitaevskii2003bose}. 
Indeed, it is already of significant recent interest to explore the interaction of  
vortical patterns, as  has been done  
in two-component settings, e.g., in~\cite{richaud1,richaud2} (see also references therein). 
Moreover, in the current setup the inclusion of three-body recombination processes 
as a dissipative mechanism in selective spin-channels constitutes a situation that accounts 
for possible experimental imperfections~\cite{huh2020observation}. 
Yet another fruitful perspective is to consider domain-walls formed by two out 
of the three spin-components with the remaining one being a nonlinear excitation of different flavor, e.g. a vortex~\cite{yu2021dark}. 
This setting will enable one to devise particular spin-mixing channels and consequently study dynamical pattern formation. 

\acknowledgments 
This work is funded by the Cluster of Excellence `Advanced Imaging of Matter’ of the Deutsche
Forschungsgemeinschaft (DFG) - EXC 2056 - project ID 390715994. 
S.I.M. gratefully acknowledges financial support from the NSF through a grant for ITAMP at Harvard University and in the framework of the Lenz-Ising Award of the University of Hamburg. 
This material is based upon work supported by the US National Science Foundation under Grant No. PHY-2110030 (P.G.K.).

\appendix 
\section{Elements of the BdG equation} \label{sec:appendix1}

In this appendix the distinct matrix elements of the BdG Eq.~(\ref{eigprob}) discussed in the main text are provided. 
In particular, the $2\times 2$ sub-matrices $M_j$ with $j=1,\ldots,9$ have the form  
\begin{equation*}
M_1=
    \left[ {\begin{array}{cc}
    A_{11} & A_{12}   \\
    -A^*_{12} & -A_{11} \\
    \end{array} } \right], \\
M_2=
    \left[ {\begin{array}{cc}
    A_{13} & A_{14}   \\
    -A^*_{14} & -A^*_{13} \\
    \end{array} } \right], \\
\end{equation*}
\begin{equation*}
M_3=
    \left[ {\begin{array}{cc}
    A_{15} & A_{16}   \\
    -A^*_{16} & -A^*_{15} \\
    \end{array} } \right], \\
M_4=
    \left[ {\begin{array}{cc}
    A^*_{13} & A_{14}   \\
    -A^*_{14} & -A_{13} \\
    \end{array} } \right], \\
\end{equation*}
\begin{equation*}
M_5=
    \left[ {\begin{array}{cc}
    A_{33} & A_{34}   \\
    -A^*_{34} & -A_{33} \\
    \end{array} } \right], \\
M_6=
    \left[ {\begin{array}{cc}
    A^*_{35} & A_{36}   \\
    -A^*_{36} & -A^*_{35} \\
    \end{array} } \right], \\
\end{equation*}
\begin{equation*}
M_7=
    \left[ {\begin{array}{cc}
    A^*_{15} & A_{16}   \\
    -A^*_{16} & -A_{15} \\
    \end{array} } \right], \\
M_8=
    \left[ {\begin{array}{cc}
    A^*_{35} & A_{36}   \\
    -A^*_{36} & -A_{35} \\
    \end{array} } \right], \\
\end{equation*}
\begin{equation}
M_9=
    \left[ {\begin{array}{cc}
    A_{55} & A_{56}   \\
    -A^*_{56} & -A_{55} \\
    \end{array} } \right].   \\
\label{ms}    
\end{equation}
The corresponding matrix elements $A_{ij}$ read
\begin{eqnarray}
A_{11}&=&\mathcal{H}-\mu_0+c_{0}(\abs{\Psi^0_{1}}^2 + 2\abs{\Psi^0_{0}}^2 + \abs{\Psi^0_{-1}}^2) \nonumber \\ 
&+& c_1(\abs{\Psi^0_{1}}^2 + \abs{\Psi^0_{-1}}^2), \nonumber \\ 
A_{33}&=&\mathcal{H}-\mu_1+q+c_{0}(2\abs{\Psi^0_{1}}^2 + \abs{\Psi^0_{0}}^2 + \abs{\Psi^0_{-1}}^2) \nonumber \\ 
&+& c_1(2\abs{\Psi^0_{1}}^2 + \abs{\Psi^0_{0}}^2 - \abs{\Psi^0_{-1}}^2), \nonumber \\
A_{55}&=&\mathcal{H}-\mu_{-1}+q+c_{0}(\abs{\Psi^0_{1}}^2 + \abs{\Psi^0_{0}}^2 + 2\abs{\Psi^0_{-1}}^2) \nonumber \\ 
&+& c_1(2\abs{\Psi^0_{-1}}^2 + \abs{\Psi^0_{0}}^2 - \abs{\Psi^0_{1}}^2), \nonumber \\
A_{12}&=&c_{0}{\Psi^0_{0}}^2 + 2c_1\Psi^0_{-1}\Psi^0_{1}, \nonumber \\
A_{13}&=&\left(c_{0}+c_1\right){\Psi^0_{1}}^*\Psi^0_{0} + 2c_1{\Psi^0_{0}}^*\Psi^0_{-1}, \nonumber \\
A_{14}&=&\left(c_{0}+c_1\right){\Psi^0_{1}}\Psi^0_{0}, \nonumber \\
A_{15}&=&\left(c_{0}+c_1\right)\Psi^{0^*}_{-1}\Psi^0_{0} + 2c_1{\Psi^0_{0}}^*\Psi^0_{1}, \nonumber \\
A_{16}&=&\left(c_{0}+c_1\right){\Psi^0_{-1}}\Psi^0_{0}, A_{34} =\left(c_{0}+c_1\right){\Psi^0_{1}}^2, \nonumber \\
A_{35}&=&\left(c_{0}-c_1\right){\Psi^{0^*}_{-1}}\Psi^0_{1}, A_{36}=\left(c_{0}-c_1\right)\Psi^{0}_{-1}\Psi^0_{1} + c_1\Psi^{0^2}_{0}, \nonumber \\
A_{56}&=&\left(c_{0}+c_1\right){\Psi^{0^2}_{-1}}. %\nonumber 
\label{alphas}
\end{eqnarray}
Recall, that $\Psi^0_{m_F}(x,y)$ denotes the 
relevant for each magnetic phase equilibrium 
solution. 
Substituting Eqs.~(\ref{ms}) and 
Eqs.~(\ref{alphas}) in the eigenvalue problem 
of Eq.~(\ref{eigprob}) leads, upon numerical 
evaluation, to the BdG spectra given in the main text.

\section{Impact of larger system sizes and higher-charge vorticity}\label{sec:appendix} 

Here, we aim to generalize our findings presented in the main text 
by considering different system sizes and vortex charges. 
In particular, in the former case we systematically vary the total number of particles within 
the range $N \in [1 \times 10^3, 2 \times 10^4]$ while in the latter situation vortices of $S=2, 3$ are explored. 
Experimentally higher-charge vortices can be realized using the topological phase-imprinting technique~\cite{leanhardt2002imprinting}. 
Remarkably enough, by monitoring the polarization of the FM spinor system under $(q, N)$ variations reveals that  
it remains insensitive under such parametric changes independently of the stationary configuration (not shown for brevity). 
Sizable deviations are only present when higher charge vortices are contained either in a VBV or a BVB equilibrium solution. 
Indeed, as presented in Fig.~\ref{fig_6}$(\rm{a})$-$(\rm{b})$, $P$ experiences drastic changes under a  $(q, S)$ variation. 
Particularly, while $S$ increases an overall shift of $P$ towards more positive (negative) $q$ values is observed for BVB (VBV) 
solutions altering in this way the distinct magnetic phase transition boundaries. 
Since $S=2,3$ vortices are structures having significantly wider cores, 
see Fig.~\ref{fig_6}$(\rm{c_1})$-$(\rm{c_3})$ and Fig.~\ref{fig_6}$(\rm{d_1})$-$(\rm{d_3})$ respectively, when compared to 
the $S=1$ configurations (see the insets in Fig.~\ref{fig_4}), the above-mentioned shift can be explained as follows. 
Initially, we should recall that bright solitons can only be sustained in repulsive environments, via their effective trapping 
by nonlinear excitations such as the vortices studied herein~\cite{pola2012vortex}. 
Thus, higher charge vortices can effectively trap in a more efficient manner the bright soliton component leading in turn to 
persistent over wider parametric intervals 3C entities.
    
\begin{figure*}
\begin{center}
\includegraphics[width=1.0\textwidth]{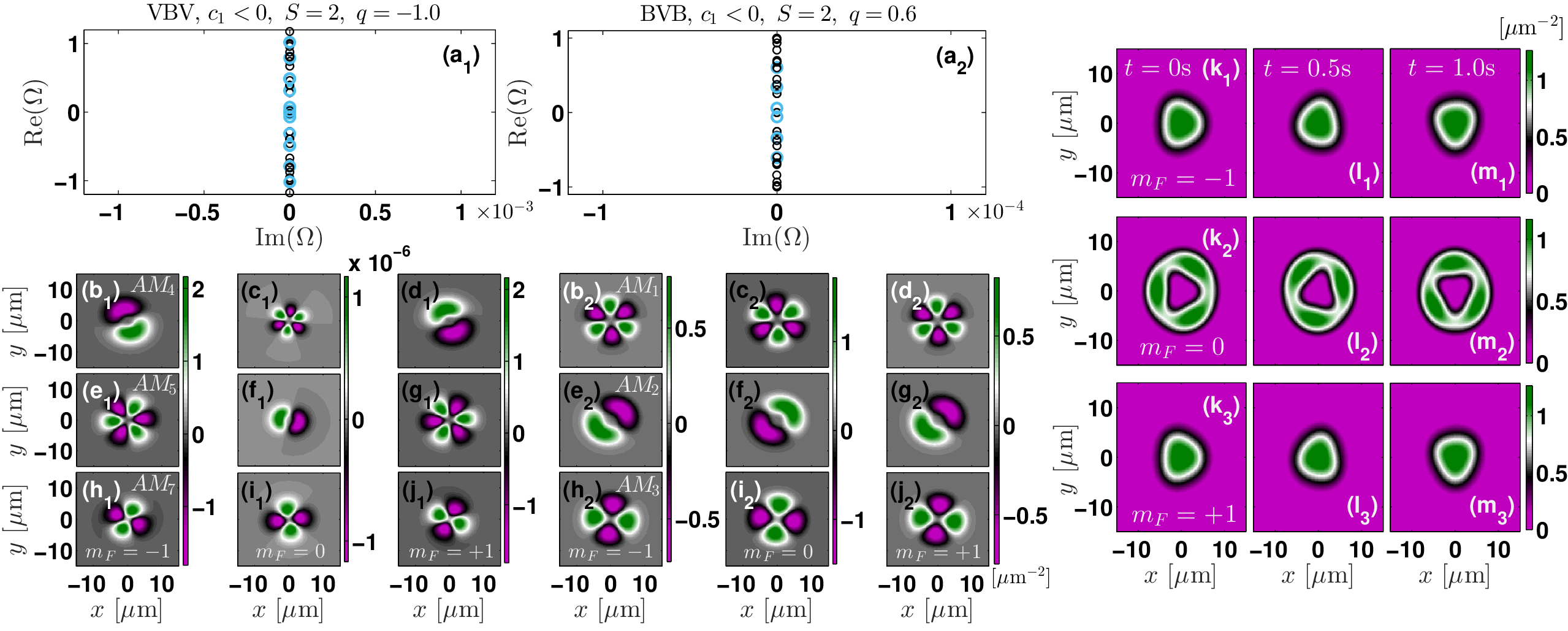}
\caption{ $(\rm{a_1})$ [$(\rm{a_2})$] BdG spectra of doubly quantized, $S=2$, VBV [BVB] 
stationary states for $q=-1.0$ [$q=0.6$] and for FM interactions ($c_1<0$). 
Notice the absence of imaginary eigenfrequencies for both entities that demonstrates their spectral stability. 
Remarkably seven $AM_i$ (with $i=1,2,\ldots,7$) pairs, being marked by light blue circles, 
are present for the VBV configuration in contrast to the three found for a BVB solution. 
$(\rm{b_1})-(\rm{j_1})$ [$(\rm{b_2})-(\rm{j_2})$] 2D contour plots measuring the density difference $\Delta \Psi_{m_F}(x,y) \equiv |\tilde{\Psi}_{m_F}(x,y)|^2-|\Psi_{m_F}(x,y)|^2$ 
for a VBV [BVB] solution and for $q=-1.0$ [$q=0.6$] (see legends). 
($\rm{k_1}$)--($\rm{m_3}$) Snapshots of the density, $|\tilde{\Psi}_{m_{F}}(x,y)|^2$, of an $S=2$ BVB solution for $q=0.6$ perturbed 
via the eigenvector associated with $AM_1$. 
The distinct spin-components are shown respectively for $t=0$s, $t=0.5$s and $t=1.0$s (see legends). 
All densities are illustrated in dimensional units.}
\label{fig_7}
\end{center}
\end{figure*}

Even though it is known that multiply-quantized vortices are prone to decay into singly quantized 
vortex pairs in scalar~\cite{pu1999coherent,leanhardt2002imprinting,mottonen2003splitting,shin2004dynamical,huhtamaki2006splitting} 
and two-component BECs~\cite{kevrekidis2015defocusing}, the fate of such higher charge entities in spinorial BEC systems remains still
elusive~\cite{leanhardt2003coreless}. 
As such, below we further investigate the stability properties of these configurations. 
Specifically, we focus on the simplest case scenario, namely the one involving spinors in which the vortices have charge $S=2$. 
Our stability analysis reveals that doubly quantized FM VBV and BVB are, in principle, 
linearly stable configurations for values of $q\in [-4.0, -0.5)$ and $q\in [0.2, 4.0]$ 
respectively that we have checked and for the particular particle number chosen. 
Narrow windows where oscillatory instabilities are identified, giving rise to a finite imaginary contribution 
of the order of $\rm{Im}(\Omega) \sim 10^{-3}-10^{-2}$, occur for the VBV configuration e.g. for $q \in [-0.75, 0.65]$ and $q=-0.9$. 
Remarkably, seven negative energy modes, $AM_i$ ($i=1,2,\ldots,7$), are found in the BdG spectrum of 
this structure as can be seen for instance in Fig.~\ref{fig_7}$(a_1)$ for $q=-1.0$. 

Among these modes the lowest-lying one, $AM_1$, resides close to the zero frequency axis, as in the FM $S=1$ scenario.
In order to visualize the effect that the perturbation has on the VBV excitation, we invoke, as in the main text, the density difference, $\Delta \Psi_{m_F}(x,y) \equiv |\tilde{\Psi}_{m_F}(x,y)|^2-|\Psi_{m_F}(x,y)|^2$. 
It turns out that contrary to the 
$S=1$ case here $\Delta \Psi_{\pm 1}(x,y)$
develops an eight {\it lobe} dumbbell-shaped structure centered around the origin of the $(x-y)-$plane ($x=y=0$) and being a reminiscent of a $5{\rm g_{z^3x}}$ orbital. 
This density difference is further found to acquire its maximum/minimum value in an alternating fashion among the distinct {\it lobes.} 
Importantly though, also for higher charges, the number of negative Krein modes is greater than the one anticipated for an $S=2$ VBV solution. 
Indeed, it is known~\cite{kevrekidis2015defocusing} that since the two vortices are doubly quantized in this case 
one can assign two anomalous mode pairs to each of the two participating vortices. 
These yield in turn four anomalous mode pairs for such a state rather than the seven identified herein. 

Thus, in what follows $\Delta \Psi_{m_F}(x,y)$ is evaluated and shown in Fig.~\ref{fig_7}$(\rm{b_1})-(\rm{j_1})$ for three out of the seven modes that VBV solutions possess. 
Notice that in all three cases the bright soliton of the $m_F=0$ spin-component is not altered as captured by $\Delta \Psi_{0}(x,y)\sim 10^{-6}$. 
This is in contrast to the vortices of the $m_F=\pm1$ spin-components that complement one another.  
Evidently, perturbing the VBV solution with the eigenvector 
related to $AM_4$ results in an asymmetric two {\it lobe}
$\Delta \Psi_{\pm 1}(x,y)$ configuration resembling a 2p orbital. The latter, is  oriented along the anti-diagonal $x=-y$ but is slightly shifted from it counterclockwise. $\Delta \Psi_{-1}(x>0,y)>0$ while $\Delta \Psi_{+1}(x>0,y)<0$.   
$AM_5$ leads to a centered around the origin $4\rm{f_{xz^2}}$ orbital-like configuration. Namely, a six dumbbell-shaped {\it lobe} structure [Fig.~\ref{fig_7}$(\rm{e_1})$ and $(\rm{g_1})$]. 
Notice that the density difference maximizes and minimizes in an alternating manner as we go from one {\it lobe} to the other.
Here, dynamical activation of $AM_4$ unveils 
the formation of anti-phase triangular 
patterns in the vortex $m_F=\pm 1$ components which along with an intact bright soliton 
$m_F=0$ component precess around the trap 
center. 
Contrary to the above dynamics, perturbing 
the VBV entity with $AM_5$ leads to the 
formation and robust propagation of a 
deformed structure. 
The two vortices perform an irregular 
out-of-phase precession leaving in this way 
the bright soliton in the $m_F=0$ component 
intact, but instead of forming triangles, 
they feature dipolarly elongated density 
distributions being inverted between the 
$m_F=+1$ and $m_F=-1$ components. 
However, addition of the eigenvector 
associated with $AM_7$ entails a completely 
different deformation. 
$\Delta \Psi_{\pm 1}(x,y)$ develops a 
$3d_{xz}$ orbital-like pattern 
[Fig.~\ref{fig_7}$(\rm{h_1})$ and 
$(\rm{j_1})$]. 
That is, a four {\it lobe} cloverleaf 
distribution with the symmetric hyperfine 
components complementing one another.

Dynamical activation of $AM_7$ leads to a 
breathing core VBV structure that performs an irregular (out-of-phase) precession having 
spatially anisotropic and oppositely elongated 
with respect to each other symmetric spin components. 
The remaining eigenvectors associated with $AM_2$, $AM_3$ and $AM_6$ result respectively in a $\Delta \Psi_{m_F}(x,y)$ that has a $4d_{xz}$ orbital structure in all three hyperfine components, having $\Delta \Psi_{0}(x,y)\sim 10^{-8}$ and being centered at the origin of the $(x-y)-$plane.  
$AM_3$ leads to a 2p orbital distribution like the one found for the $AM_4$ mode but with the two {\it lobes} being slightly shifted with respect to each other while residing anti-diametrically along the diagonal $x=y$. 
Here, $\Delta \Psi_{-1}(x,y)=\Delta \Psi_{+1}(x,y)$ and both are complementary to the $m_F=0$ bright soliton component.
Additionally, the effect of $AM_6$ closely resembles that found for $AM_7$ but with the symmetric vortex components having now exactly the same structure while being complementary to $\Delta \Psi_{0}(x,y)$ which is now finite. 
Finally, we note that $AM_1$ and $AM_2$ perform an eigenfrequency zero 
crossing at $q=-0.4$ but are not responsible for an instability (Im$(\Omega=0)$). 
The rest of the $AM$, i.e. $AM_{i}$ with $i=3,\ldots, 7$, decrease in frequency but 
only around $q=-0.05$ cross the zero frequency axis signaling the termination of this nonlinear excitation.

On the other hand, $S=2$ BVB solutions destabilize via two eigenfrequency zero crossings of the two principal AMs 
present in the BdG spectrum of this configuration. 
Namely, $AM_3$ which is the higher-lying negative energy mode and $AM_2$ being the lowest-lying one. 
These destabilizations take place at $q=0$, i.e., at the threshold ($q=0$) separating the EP and the EA phases, and $q=0.15$. 
However, among the two only the second destabilization produces a sizable imaginary component 
being of the order of $\rm{Im}(\Omega) \sim 10^{-2}$. 
Also an oscillatory instability is identified for the $S=2$ BVB entity appearing at around $q=0.7$. 
This is an instability that owes its existence to the collision of $AM_3$ with a positive Krein mode giving rise to 
an eigenfrequency quartet similar to those identified for the $S=1$ structures. 
There exists also a third anomalous mode for this BVB configuration. 
Namely $AM_1$, that stems from a change in sign of a background mode from positive to negative. 
This mode appears in the BdG spectrum for $q=0.6$ and remains present as $q$ is further lowered towards the phase transition point. 

The above-discussed modes are illustrated in 
Fig.~\ref{fig_7}$(\rm{a_2})$ while their 
activation leads to deformations of the 
stationary $S=2$ BVB state, an effect 
that is measured via $\Delta \Psi_{m_F}(x,y)$ 
shown in 
Fig.~\ref{fig_7}$(\rm{b_2})-(\rm{j_2})$. 
Notice that $\Delta \Psi_{m_F}(x,y)$
is finite irrespectively of which mode, i.e. $AM_1$, $AM_2$ and $AM_3$, is activated.
Particularly, for the first mode at hand, $\Delta \Psi_{m_F}(x,y)$ acquires a $4{\rm f_{xz^2}}$ orbital-like distribution as the 
one found for the perturbed via $AM_5$ VBV entity. Here though, $\Delta \Psi_{-1}(x,y)=\Delta \Psi_{+1}(x,y)$ while both are complementary to the $m_F=0$ vortex component.
Likewise, the density difference assumes a 2p orbital-like structure once $AM_2$ is taken into account, a result similar to the one found for the VBV solutions when $AM_3$ was triggered. 

Finally, the eigenvector related to $AM_3$ is responsible for a $3d_{xz}$ deformation imprinted in $\Delta \Psi_{m_F}(x,y)$ like the one found for the VBV structure when perturbed 
with the eigenvector associated with $AM_7$ [see here Fig.~\ref{fig_7}$(\rm{h_2})$- $(\rm{j_2})$. 
However here, $\Delta \Psi_{-1}(x,y)=\Delta \Psi_{+1}(x,y)$ while both are complementary to $\Delta \Psi_{0}(x,y)$ that is also finite in this case.

A case example showcasing the dynamical 
evolution of a perturbed $S=2$ configuration 
is provided in Fig.~\ref{fig_7}$(\rm{k_1})-(\rm{m_3})$ 
for $q=0.6$. 
Notice the structural deformation of the 
ensuing BVB structure caused by the addition 
of the eigenvector related to $AM_1$. 
Evidently, already at $t=0$s a triangular 
pattern~\cite{maity2020parametrically,PhysRevLett.127.113001,saint2019dynamical}, 
breaking the radial symmetry of the trap 
along the azimuthal direction, is seen 
in 
Fig.~\ref{fig_7}$(\rm{k_1})-(\rm{k_3})$ 
whose precessional motion is then 
followed for times up to $t=1.0$s 
[Fig.~\ref{fig_7}$(\rm{m_1})-(\rm{m_3})$]. 
An outcome verifying that indeed,
this deformation is caused by the 
above-identified azimuthal mode 
with triangular symmetry 
(i.e., an $e^{3i \theta}$ perturbation 
mode). 
It is also worthwhile to mention that similar findings are also present for AF spinor BECs 
(not shown).

\bibliographystyle{apsrev4-1}
\bibliography{references}
\end{document}